\documentclass[aps,pre,twocolumn]{revtex4}
\usepackage{graphics}
\usepackage{graphicx}

\begin{document}


\title{A topological approximation of the nonlinear Anderson model}


\author{Alexander~V.~Milovanov}

\affiliation{ENEA National Laboratory, Centro~Ricerche~Frascati, I-00044 Frascati, Rome, Italy}
\affiliation{Space Research Institute, Russian Academy of Sciences, 117997 Moscow, Russia}
\affiliation{Max-Planck-Institut f\"ur Physik komplexer Systeme, 01187 Dresden, Germany}

\author{Alexander~Iomin}
\affiliation{Department of Physics and Solid State Institute, Technion, Haifa, 32000, Israel}
\affiliation{Max-Planck-Institut f\"ur Physik komplexer Systeme, 01187 Dresden, Germany}



\begin{abstract} 
We study the phenomena of Anderson localization in the presence of nonlinear interaction on a lattice. A class of nonlinear Schr\"odinger models with arbitrary power nonlinearity is analyzed. We conceive the various regimes of behavior, depending on the topology of resonance-overlap in phase space, ranging from a fully developed chaos involving L\'evy flights to pseudochaotic dynamics at the onset of delocalization. It is demonstrated that the quadratic nonlinearity plays a dynamically very distinguished role in that it is the only type of power nonlinearity permitting an abrupt localization-delocalization transition with unlimited spreading already at the delocalization border. We describe this localization-delocalization transition as a percolation transition on the infinite Cayley tree (Bethe lattice). It is found in vicinity of the criticality that the spreading of the wave field is subdiffusive in the limit $t\rightarrow+\infty$. The second moment of the associated probability distribution grows with time as a powerlaw $\propto t^\alpha$, with the exponent $\alpha = 1/3$ exactly. Also we find for superquadratic nonlinearity that the analog pseudochaotic regime at the edge of chaos is self-controlling in that it has feedback on the topology of the structure on which the transport processes concentrate. Then the system automatically (without tuning of parameters) develops its percolation point. We classify this type of behavior in terms of self-organized criticality (SOC) dynamics in Hilbert space. For subquadratic nonlinearities, the behavior is shown to be sensitive to details of definition of the nonlinear term. A transport model is proposed based on modified nonlinearity, using the idea of ``stripes" propagating the wave process to large distances. Theoretical investigations, presented here, are the basis for consistency analysis of the different localization-delocalization patterns in systems with many coupled degrees of freedom in association with the asymptotic properties of the transport. 
\end{abstract}


\maketitle

\section{Introduction} The dynamics of systems with competition between dispersion, randomness, and nonlinearity constitutes a problem of universal significance. The various aspects of it are encountered in for instance nonlinear Schr\"odinger models with dispersive interactions \cite{Christ,Gaid}, expansion of a Bose-Einstein condensate in the presence of disorder \cite{Clement,Shapiro}, fracton-mediated superconductivity in complex superconductors \cite{Blumen,PRB02}, localization phenomena in Fock space and the problem of electron-electron lifetime in a quantum dot \cite{Alt}. It has been proposed recently \cite{Chapter} that the formalism of discrete Anderson nonlinear Schr\"odinger equation with self-adjusting nonlinearity offers a conceptual framework for the phenomena of self-organized criticality (SOC) \cite{Bak} governing the natural occurrence of attractive critical states in nonlinear driven systems.   

In this study, we consider the problem of Anderson localization of waves in a class of nonlinear Schr\"odinger models with random potential on a lattice and arbitrary power nonlinearity. The phenomena of Anderson localization are based on interference between multiple scattering paths, leading to localized wavefunctions with exponentially decaying profiles \cite{And}. Experimentally, Anderson localization has been reported for electron gases \cite{Electron}, acoustic waves \cite{Sound}, light waves \cite{Maret,Light}, and matter waves in a controlled disorder \cite{BE}. 

{\it The challenge} $-$ A prospective new feature, arising in these phenomena, is destruction of Anderson localization by a weak nonlinearity, as computer simulations in discrete geometry show \cite{Sh93,PS,Flach,Skokos}. 
Indeed it was argued based on numerical modeling that above a certain critical strength of the nonlinear interaction there is a delocalization border \cite{PS,EPL}, beyond which the field spreads unlimitedly along the lattice, and is dynamically localized despite these nonlinearities otherwise. 
A good candidate to investigate and explain the phenomenon from first principles is at present the Gross-Pitaevskii equation \cite{Gross}, also known as the nonlinear Schr\"odinger equation. It has been rigorously established, for a large variety of interactions and of physical conditions, that the Gross-Pitaevskii equation is exact in the thermodynamic limit \cite{Edros}. Theoretically, nonlinear Schr\"odinger models offer a mean-field approximation, where the term containing the probability density absorbs the interactions between the components of the wave function. 

Despite these advances, a detailed understanding of the mechanisms driving the delocalization is still not at hand. In particular, it is not clear what is the long time ($t\rightarrow+\infty$) asymptotic behavior of an initially localized wave packet, if {\it both} nonlinearity and randomness are present on an equal footing. Another important issue here concerns the type of the nonlinearity permitting delocalization. The goal of the present study is to obtain progress over these topics.
 
{\it The method} $-$ We propose a systematic approach to the phenomena of dynamical localization-delocalization in random media through a topological approximation of the nonlinear Anderson model. We analyze delocalization processes as a transport problem for a dynamical system with many coupled degrees of freedom, with an emphasis on the criticality aspects of delocalization. A short account of this approach has been reported previously \cite{EPL}. 

Mathematically, our work is based on the general studies of transition to chaos in Hamiltonian systems \cite{Sagdeev,Report}. We conceive the various patterns of behavior, depending on the topology of resonance-overlap in phase space, ranging from a fully developed chaos involving L\'evy flights \cite{Report,Klafter,Chechkin} to regular (Kolmogorov-Arnold-Moser, or KAM, regime) dynamics \cite{Arnold,ChV}. A borderline regime separating the chaotic and the regular cases is also discussed, and is associated with Hamiltonian pseudochaos, i.e., random nonchaotic dynamics with zero Lyapunov exponents \cite{Report,JMPB,PhysD}. 

{\it Schedule and outlook} $-$ In what follows, we shall first consider a nonlinear Schr\"odinger model with quadratic nonlinearity. Then we shall cast the model in a more general context of arbitrary power nonlinearity and raise some questions suggested by this generalization. This schedule of the presentation is motivated by the extraordinary important implications of quadratic nonlinearity in the phenomena of chaotic transport in phase space \cite{Sagdeev,ChV}, in borderline behavior, and in critical localization-delocalization phenomena. 

Indeed we find using a topological mapping procedure that quadratic nonlinearity plays a dynamically very distinguished role in that it is the only type of power nonlinearity permitting an abrupt localization-delocalization transition with unlimited spreading of the wave function already at the delocalization border. That means that Anderson localization can survive only a finite strength of nonlinearity of the quadratic type, and that the destruction of localization is a phase transition-like phenomenon. 
We describe this localization-delocalization transition as a percolation transition on the infinite Cayley tree (Bethe lattice). 

Focusing on the quadratic power nonlinearity, we find in vicinity of the criticality that the spreading of the wave field is {subdiffusive}: The second moment of the associated probability distribution grows with time as a powerlaw $\propto t^{\alpha}$ for $t\rightarrow+\infty$, 
with the exponent $\alpha = 1/3$ exactly. We should emphasize that the observed subdiffusive behavior is {\it asymptotic}. 
This critical regime is very special in that it stems from the direct proportionality between the nonlinear frequency shift and the distance between the excited modes in wave number space. 
Topologically, the phenomena of critical spreading correspond with a next-neighbor random walk at the onset of percolation on a Cayley tree. 

Extending this borderline behavior to superquadratic nonlinearities is not at all trivial. It leads to a self-controlling transport with feedback of the spreading process on the dynamical state of the lattice. We demonstrate that the system self-adjusts while spreading to stabilize exactly at the point of critical percolation in Hilbert mapping space; where it is characterized by marginal topological connectedness with the infinitely remote point \cite{Chapter,PRE97}. Thus, threshold percolation attracts the feedback spreading, implying that the phase space of the system is by itself a dynamic medium coupled with the transport of the waves. We classify this type of behavior in terms of an SOC dynamics in the Hilbert space. 
The asymptotic state of the wave field is found at the border of regularity with virtually no transport in the limit $t\rightarrow+\infty$. 

The phenomena of critical spreading find their significance in association with the general problem of transport along separatrices of dynamical systems with many coupled degrees of freedom \cite{Arnold,ChV}. Mathematically, they correspond to a long-time correlated behavior at the edge of chaos (with or without a feedback, depending on the type of the nonlinearity that is analyzed). In the quadratic nonlinearity case, the spreading process being unlimited allows for a statistical description \cite{Chapter,PRE09,NJP} in terms of fractional diffusion equation with the fractional derivative in time (e.g., Refs. \cite{Report,Klafter} for reviews), also demonstrating an important interconnection between fractional kinetics and Hamiltonian pseudochaos \cite{Report}.

With the departure from borderline behavior, the nonlinear properties take a stronger role over the dynamics, giving rise to a pure chaotic behavior of the Fokker-Planck type. This crossover to chaos \cite{Sagdeev,ZaslavskyUFN} is confirmed for both quadratic and superquadratic nonlinearities. Even so, for superquadratic nonlinearity, there is no a ``universal" transition point to chaos (nor {unlimited} spreading at the delocalization border $-$ at contrast with the quadratic nonlinearity case) in that the crossover involves, as a general situation, the number of already occupied states and, therefore, is dynamic.  

Indeed our results show that (i) there {\it always} exists a critical strength of the nonlinearity parameter separating the chaotic and the regular transport regimes; (ii) this critical strength is only {\it preserved} through dynamics for quadratic nonlinearity; and (iii) is dynamically {\it evolving} through the dependence on the number of already occupied states otherwise. The differentiation between the chaotic and the pseudochaotic regimes is very important in this description, as it helps to sort out some ambiguities in the reported transport exponents \cite{PS,Flach}; as well as to place the various transport models on a solid mathematical background in connection with the asymptotic character of the transport.

In the chaotic regime, we find for quadratic nonlinearity a subdiffusion of waves with the exponent $\alpha = 2/5$. This subdiffusion occurs as a consequence of range-dependence of the diffusion coefficient, designated by the cubic interaction between the components of the wave field, and has Markovian (memoryless) character. The analog behavior in the case of superquadratic nonlinearity exists, and is well defined. It leads to yet a slower spreading of the subdiffusive type, with the $\alpha$ value inversely proportional with the power nonlinearity. Essentially the same scaling is obtained for the front of diffusion on loopless fractals; where a minimal-distance the so-called chemical metric \cite{Havlin} is defined by the overlap integral in the nonlinear Anderson model. 

That the chaotic dynamics are Fokker-Planck is a standard paradigm based on the central limit theorem of the theory of the probability, using finiteness of second and higher moments of a sum of independent random variables \cite{Van}. Generalizations of this correspond to the L\'evy-Gnedenko central limit theorem dealing with random processes with diverging variance \cite{Klafter,Gnedenko}. This generalized central limit theorem when applied to dynamical delocalization problem leads to a theoretical possibility of field-spreading by L\'evy flights. This regime is remarkable, as the dynamics are controlled by a competition between the nonlocality of the L\'evy motion \cite{Klafter,Chechkin} and the nonhomogeneity of the nonlinear interaction associated with a range-dependent driving noise process. 

The origin of L\'evy motion in random media can be attributed \cite{PLA} to a competing nonlocal ordering \cite{PRB02,Cho}; which is {\it subordinate} to the dynamical ordering due to for instance a nonlinear wave field. That means that the dynamical order parameter acts as input control parameter for the nonlocal order \cite{Chapter,UFN}; thus giving rise to a statistics of the L\'evy type through self-organization \cite{JJR}. Then the nonlinear wave can generate via a back-reaction on the medium a channel along which it propagates to large distances on L\'evy flights \cite{UFN}. One example of this behavior is the self-generation of so-called ``stripy" ordering, which is discussed in Sec. III, part E, based on one-dimensional percolation model, using a nonlinear Schr\"odinger equation with modified nonlinearity. 

Most interestingly, we find that the inclusion of L\'evy flights does not really destroy the subdiffusive character of field-spreading, so the transport is {\it simultaneously} subdiffusive and nonlocal. This is because the nonlinear interaction term introduces strong range-dependence into the intensity of the L\'evy noise; which dies away at a fast pace while spreading. Then there is an upper bound on the rate of nonlocal transport, and this corresponds to a diffusive scaling with $\alpha = 1$. 
We hasten to note that this ``diffusion" of the wave function is absolutely anomalous in that it arises from a compromise between the nonlocality of L\'evy flights and the topological constraints stipulated by the nonlinear interaction term.  

For subquadratic nonlinearities, the behavior is shown to be sensitive to details of definition of the nonlinear term. A transport model is proposed based on modified nonlinearity, incorporating the idea of ``stripes" propagating the wave process to large distances. This type of transport is only possible in the chaotic regime. 
In the absence of stripes, the nonlinear field is localized in much the same way as linear field. In those cases, the nonlinearity merely shifts the energy levels across the system without destroying the overall localized state. 

Finally, we suggest using the modified model that the stripy ordering and the associated destruction of Anderson localization by an algebraic nonlinearity account for the ``paradoxical" existence of superconductivity in some disordered superconductors \cite{Cho,Emery}; 
where the usual linear theories predict localization of the superconducting wave function by the underlying molecular disorder.   


{\it Before we start off} $-$ We would like to emphasize that the transport regimes, which we discuss, are asymptotic regimes in the limit $t\rightarrow+\infty$. Because of this asymptotic character, their validation through computer simulations is an important yet intricate task, if only due to coarse-graining of parameters defining a system at criticality as well as the natural limitations with respect to finite size effects and the possible lack of the statistics. The theoretical approaches, presented here, are aimed to provide the basis for consistency analysis of the various transport regimes in nonlinear Schr\"odinger models with disorder for arbitrary power nonlinearity.  

\section{Quadratic nonlinearity} We work with a variant of discrete Anderson nonlinear Schr\"odinger equation (DANSE)
\begin{equation}
i\hbar\frac{\partial\psi_n}{\partial t} = \hat{H}_L\psi_n + \beta |\psi_n|^2 \psi_n,
\label{1} 
\end{equation}
with
\begin{equation}
\hat{H}_L\psi_n = \varepsilon_n\psi_n + V (\psi_{n+1} + \psi_{n-1}).
\label{2} 
\end{equation}
Here, $\hat H_L$ is the Hamiltonian of a linear problem in the tight binding approximation; $\beta$ ($\beta > 0$) characterizes the strength of nonlinearity; on-site energies $\varepsilon_n$ are randomly distributed with zero mean across a finite energy range; $V$ is hopping matrix element; and the total probability is normalized to $\sum_n |\psi_n|^2 = 1$. In what follows, $\hbar = 1$ for simplicity. In the absence of randomness, DANSE~(\ref{1}) is completely integrable. For $\beta\rightarrow 0$, the model in Eqs.~(\ref{1}) and~(\ref{2}) reduces to the original Anderson model in Refs. \cite{And,And+}. All eigenstates are exponentially localized in this limit with dense eigenspectrum. We aim to understand the asymptotic ($t\rightarrow+\infty$) spreading of initially localized wave packet under the action of nonlinear term.    

Expanding $\psi_n$ over a basis of linearly localized modes, the eigenfunctions of the linear problem, $\{\phi_{n,m}\}$, $m= 1,2,\dots$, we write, with time depending complex coefficients $\sigma_m (t)$,
\begin{equation}
\psi_n = \sum_m \sigma_m (t) \phi_{n,m}.
\label{3} 
\end{equation}
We consider $\psi_n$, $\psi_n\in\{\psi_n\}$, as a vector in functional space, whose basis vectors $\phi_{n,m}$ are the Anderson eigenstates. For strong disorder, dimensionality of this space is infinite (countable). It is convenient to think of each node $n$ as comprising a countable number of ``compactified" dimensions representing the components of the wave field. So these hidden dimensions when account is taken for Eq.~(\ref{3}) are ``expanded" via a topological mapping procedure to form the functional space $\{\psi_n\}$. We consider this space as providing the embedding space for dynamics. Further, given any two vectors $\psi_n\in\{\psi_n\}$ and $\phi_n\in\{\psi_n\}$, we define the inner product, $\langle\psi_{n} \circ \varphi_{n}\rangle$, 
\begin{equation}
\langle\psi_{n} \circ \varphi_{n}\rangle = \sum_n \psi_n^* \varphi _{n},
\label{Inn} 
\end{equation}
where star denotes complex conjugate. To this end, the functional space $\{\psi_n\}$ becomes a Hilbert space, permitting the notions of length, angle, and orthogonality by standard methods \cite{Hirsch}. With these implications in mind, we consider the functions $\phi_{n,m}$ as ``orthogonal" basis vectors obeying
\begin{equation}
\sum _n \phi^*_{n,m}\phi_{n,k} = \delta_{m,k},
\label{Kro} 
\end{equation}
where $\delta_{m,k}$ is Kronecker's delta. Then the total probability being equal to 1 implies 
\begin{equation}
\langle\psi_n\circ\psi_n\rangle = \sum_n \psi_n^*\psi_n = \sum_m \sigma_m^* (t)\sigma_m (t) = 1.
\label{Tot} 
\end{equation}
We now obtain a set of dynamical equations for $\sigma_m (t)$. For this, substitute Eq.~(\ref{3}) into DANSE~(\ref{1}), then multiply the both sides by $\phi^*_{n,k}$, and sum over $n$, utilizing the orthogonality. The result reads  
\begin{equation}
i\dot{\sigma}_k - \omega_k \sigma_k = \beta \sum_{m_1, m_2, m_3} V_{k, m_1, m_2, m_3} \sigma_{m_1} \sigma^*_{m_2} \sigma_{m_3},
\label{4} 
\end{equation}
where $\omega_k$, $k=1,2,\dots$, are eigenvalues of the linear problem, i.e., $\hat H_L \phi_{n,k} = \omega_k \phi_{n,k}$, the coefficients $V_{k, m_1, m_2, m_3}$ are given by
\begin{equation}
V_{k, m_1, m_2, m_3} = \sum_{n} \phi^*_{n,k}\phi_{n,m_1}\phi^*_{n,m_2}\phi_{n,m_3},
\label{5} 
\end{equation}
and we have used dot to denote time differentiation. Equations~(\ref{4}) correspond to a system of coupled nonlinear oscillators with the Hamiltonian 
\begin{equation}
\hat H = \hat H_{0} + \hat H_{\rm int}, \ \ \ \hat H_0 = \sum_k \omega_k \sigma^*_k \sigma_k,
\label{6} 
\end{equation}
\begin{equation}
\hat H_{\rm int} = \frac{\beta}{2} \sum_{k, m_1, m_2, m_3} V_{k, m_1, m_2, m_3} \sigma^*_k \sigma_{m_1} \sigma^*_{m_2} \sigma_{m_3}.
\label{6+} 
\end{equation}
Here, $\hat H_{0}$ is the Hamiltonian of non-interacting harmonic oscillators and $\hat H_{\rm int}$ is the interaction Hamiltonian. Note that we have included self-interactions into $\hat H_{\rm int}$. Each nonlinear oscillator with the Hamiltonian   
\begin{equation}
\hat h_{k} = \omega_k \sigma^*_k \sigma_k + \frac{\beta}{2} V_{k, k, k, k} \sigma^*_k \sigma_{k} \sigma^*_{k} \sigma_{k}
\label{6+h} 
\end{equation}
and the equation of motion 
\begin{equation}
i\dot{\sigma}_k - \omega_k \sigma_k - \beta V_{k, k, k, k} \sigma_{k} \sigma^*_{k} \sigma_{k} = 0
\label{eq} 
\end{equation}
represents one nonlinear eigenstate in the system $-$ identified by its wave number $k$, unperturbed frequency $\omega_k$, and nonlinear frequency shift $\Delta \omega_{k} = \beta V_{k, k, k, k} \sigma_{k} \sigma^*_{k}$. Non-diagonal elements $V_{k, m_1, m_2, m_3}$ characterize couplings between each four eigenstates with wave numbers $k$, $m_1$, $m_2$, and $m_3$. It is understood that the excitation of each eigenstate is nothing else than the spreading of the wave field in wave number space. Resonances occur between the eigenfrequencies $\omega_k$ and the frequencies posed by the nonlinear interaction terms. We have
\begin{equation}
\omega_k = \omega_{m_1} - \omega_{m_2} + \omega_{m_3}.
\label{Res} 
\end{equation}
Conditions for nonlinear resonance are readily obtained by accounting for the nonlinear frequency shift. 

\subsection{Assessing the type of dynamics} When the resonances happen to overlap, a phase trajectory may occasionally switch from one resonance to another. As Chirikov realized \cite{Chirikov}, any overlap of resonances will introduce a random element to the dynamics along with some transport in phase space. Applying this argument to DANSE~(\ref{1}), one sees that destruction of Anderson localization is limited to a set of resonances in the Hamiltonian system of coupled nonlinear oscillators, Eqs.~(\ref{6}) and~(\ref{6+}), permitting a connected escape path to infinity. 

At this point, the focus is on {\it topology} of the random motions in phase space. We address an idealized situation first, where the overlapping resonances densely fill the entire phase space. This is fully developed chaos, a regime that has been widely studied and discussed in the literature (e.g., Refs. \cite{Sagdeev,Report,ZaslavskyUFN}). Concerns raised over this regime when applied to Eqs.~(\ref{6}) and~(\ref{6+}), however, come from the fact that it requires a diverging free energy reservoir \cite{EPL} in systems with a large number of interacting degrees of freedom. Even so, developed chaos offers a simple toy-model for the transport as it corresponds with a well-understood, diffusive behavior \cite{Van}.

A more general, as well as more intricate, situation occurs when the random motions coexist along with regular (KAM regime) dynamics. If one takes this idea to its extreme limit, one ends up with the general problem of transport along separatrices of dynamical systems. This problem constitutes a fascinating nonlinear problem that has as much appeal to the mathematician as to the physicist. An original important promotion of this problem to large systems is due to Chirikov and Vecheslavov \cite{ChV}.

This type of problem occurs for low frequencies \cite{PRE01}. Typically, in large systems, the set of separatrices is geometrically very complex and strongly shaped. Often it can be envisaged as a fractal network at percolation as for instance in random fields with sign-symmetry \cite{PRE00}. If the eigenfrequencies $\omega_k$ of the Hamiltonian variation are very slow, the conditions for a resonance are satisfied in vicinity of the percolating line; whose diverging length implies the vanishing of orbital frequencies, matching $\omega_k\rightarrow 0$. Indeed the percolating line is the channel through which the transport processes penetrate to the large scales \cite{EPL,PRE09,Isi}.  

It is noted that, for $\omega_k\rightarrow 0$, the resonances strongly overlap in a very narrow layer containing the percolating separatrix line; where the Chirikov's overlap condition is satisfied with a large margin \cite{PRE09,PRE01}. If one introduces for convenience the characteristic frequency of the Hamiltonian variation, $\omega$, one finds in vicinity of the separatrix that the width of the resonance layer behaves as $\Delta\omega\propto\sqrt{\omega}$; whereas the distance between the resonances approaches zero as $\delta\omega\propto\omega$. So, $\Delta\omega\gg\delta\omega$ for $\omega\rightarrow 0$. In the meanwhile, the density of resonances near the separatrix, evaluated as the inverse distance $\delta\omega$, diverges as $1/\omega$. Inside the layer, the dynamics are essentially random because of the many overlapping resonances present. The vanishing $\Delta\omega\propto\sqrt{\omega}\rightarrow 0$ implies that the random motions are squeezed into very narrow resonance layers bounded by the domains with regular behavior. Then initially close trajectories inside the layer will deviate anomalously slowly (sub-exponentially), since there is virtually no room for them to separate. The phenomenon can be described as ``stickiness" to the percolating line and is associated with the vanishing Lyapunov exponents in the layer \cite{PRE09}. %

There is a fundamental difference between the above two transport regimes (chaotic {\it vs.} near-separatrix). The former regime is associated with an exponential loss of correlation permitting a Fokker-Planck description in the limit $t\rightarrow+\infty$. The latter regime when considered for large systems is associated with an algebraic loss of correlation instead, implying that the correlation time is infinite. There is no a conventional Fokker-Planck equation here, unless extended to fractional differentiation over the time variable \cite{Report,Klafter}; nor the familiar Markovian property (i.e., that the dynamics are memoryless). On the contrary, there is an interesting interplay \cite{Chapter,PRE09} between randomness, fractality, and correlation; which is manifest in the fact that all Lyapunov exponents vanish in the thermodynamic limit, despite that the dynamics are intrinsically random. 

This situation of random non-chaotic dynamics with zero Lyapunov exponents, being indeed very general \cite{Chapter,Report}, has come to be known as ``pseudochaos." One might think of pseudochaos as occurring ``at the edge" of stochasticity and chaos, thus separating fully developed chaos from domains with regular motions. To roughly estimate the type of dynamics (regular, chaotic, or pseudochaotic) one invokes other than the number of overlapping resonances, $\Delta \mathcal{N} \gg 1$, the so-called Kubo number $\mathcal{Q}$ \cite{Isi,Kubo}, which compares frequencies of the Hamiltonian variation with those characterizing the orbital motion in phase space. Based on the results of Refs. \cite{PRE09,PRE01}, we order $\mathcal{Q} \sim (\Delta \mathcal{N})^2$ for the chaotic behavior in a wide stochastic sea; and $\mathcal{Q} \gg (\Delta \mathcal{N})^2$ for the pseudochaotic behavior near separatrix. We should stress that the number of overlapping resonances alone does {\it not} unambiguously define the type of dynamics yet, and the $\mathcal{Q}$ value is, in fact, necessary to assess the separation of trajectories. In what follows, we discuss the implications of chaotic and pseudochaotic transport for the spreading of the wave function in Eqs.~(\ref{4}).      
  
\subsection{Chaotic case} As time correlations vanish exponentially fast, Eq.~(\ref{4}) can conveniently be considered as a Langevin equation with the nonlinear interaction term thought as a Gaussian white noise term in the limit $t\rightarrow+\infty$. There is a well-defined diffusion coefficient here, which we shall denote by $D$, and which behaves as modulus squared of the intensity of the noise. The cubic interaction in Eq.~(\ref{4}) implies that $D\propto |\sigma_n| ^6$. If the field is spread over $\Delta n$ sites, then the conservation of the probability dictates $|\sigma_n| ^2 \sim 1 / \Delta n$, leading to $D\propto 1 / (\Delta n) ^3$. Hence, the transport problem in the chaotic case is basically a diffusion problem with range-dependent diffusion coefficient. This range-dependence designates the nonlinear interaction between the components of the wave field; it also introduces spatial inhomogeneity into the transport model in association with the condition that the field is initially localized. 

Further, let $f = f (t, \Delta n)$ be the probability density to find a wave packet at time $t$ at distance $\Delta n$ from the initial localization point. The diffusive character of the spreading justifies the following kinetic equation for the transport:
\begin{equation}
\frac{\partial}{\partial t} f (t, \Delta n) = \frac{\partial}{\partial \Delta n}\left[W \frac{1}{(\Delta n)^{3}}\frac{\partial}{\partial \Delta n} f (t, \Delta n) \right],
\label{7} 
\end{equation}
where $W$ is the diffusion constant and collects in a single value all parameters of the diffusion process. The fundamental solution or Green's function of Eq.~(\ref{7}) reads
\begin{equation}
f (t, \Delta n) = \frac{1/\Gamma (6/5)}{(25W)^{1/5}} t^{-1/5} \exp\left[-\frac{(\Delta n)^5}{25Wt}\right],
\label{8} 
\end{equation}
where $\Gamma$ denotes the Euler gamma-function and we have used the normalization $\int_0^\infty f (t, \Delta n) d\Delta n = 1$. Note that the distribution in Eq.~(\ref{8}) is essentially non-Gaussian as a consequence of the range-dependence of the diffusion coefficient. One sees that point-wise Gaussianity of the driving noise term does not guarantee Gaussianity of the transport process in a domain. From Eq.~(\ref{8}) one immediately obtains 
\begin{equation}
\langle (\Delta n) ^2 (t) \rangle = \left[\Gamma (3/5) / \Gamma (1/5)\right] \left[25W\right]^{2/5}t^{2/5} ,
\label{8+} 
\end{equation}
where the angle brackets denote ensemble average. The net result is $\langle (\Delta n) ^2 (t) \rangle \propto t^{2/5}$, consistently with the scaling analysis of Refs. \cite{Sh93,PS}.

\subsection{Pseudochaotic case} This regimes takes Eqs.~(\ref{4}) to the opposite extreme limit where each oscillator can only communicate with the rest of the wave field via a nearest-neighbor rule. This is a marginal regime yet permitting an escape path to infinity. Clearly, the number of coupling links is minimized in that case. When summing on the right-hand side, the only combinations to be kept are, for the reasons of symmetry, $\sigma_{k} \sigma^*_{k} \sigma_{k}$ and $\sigma_{k-1} \sigma^*_{k} \sigma_{k+1}$. We have
\begin{equation}
i\dot{\sigma}_k - \omega_k \sigma_k = \beta V_{k} \sigma_{k} \sigma^*_{k} \sigma_{k} + 2\beta V_k^\pm \sigma_{k-1} \sigma^*_{k} \sigma_{k+1},
\label{9} 
\end{equation} 
where we have also denoted for simplicity $V_k = V_{k, k, k, k}$ and $V_k^\pm = V_{k, k-1, k, k+1}$. Equations~(\ref{9}) define an infinite ($k= 1,2,\dots$) chain of coupled nonlinear oscillators where all couplings are local (nearest-neighbor-like). The interaction Hamiltonian in Eq.~(\ref{6+}) is simplified to   
\begin{equation}
\hat H_{\rm int} = \frac{\beta}{2} \sum_{k} V_{k} \sigma^*_k \sigma_{k} \sigma^*_{k} \sigma_{k} + {\beta}\sum_{k} V_k^\pm \sigma^*_k \sigma_{k-1} \sigma^*_{k} \sigma_{k+1}.
\label{6++} 
\end{equation}
We are now in position to introduce a simple lattice model for the transport. The key step is to observe that Eqs.~(\ref{9}) can be mapped on a Cayley tree where each node is connected to $z=3$ neighbors (here, $z$ is the coordination number). The mapping is defined as follows. A node with the coordinate $k$ represents a nonlinear eigenstate, or nonlinear oscillator with the equation of motion~(\ref{eq}). There are exactly $z=3$ bonds at each node: one that we consider ingoing represents the complex amplitude $\sigma^*_{k}$, and the other two, the outgoing bonds, represent the complex amplitudes $\sigma_{k-1}$ and $\sigma_{k+1}$ respectively. These settings are schematically illustrated in Fig.~1.  

\begin{figure}
\includegraphics[width=0.51\textwidth]{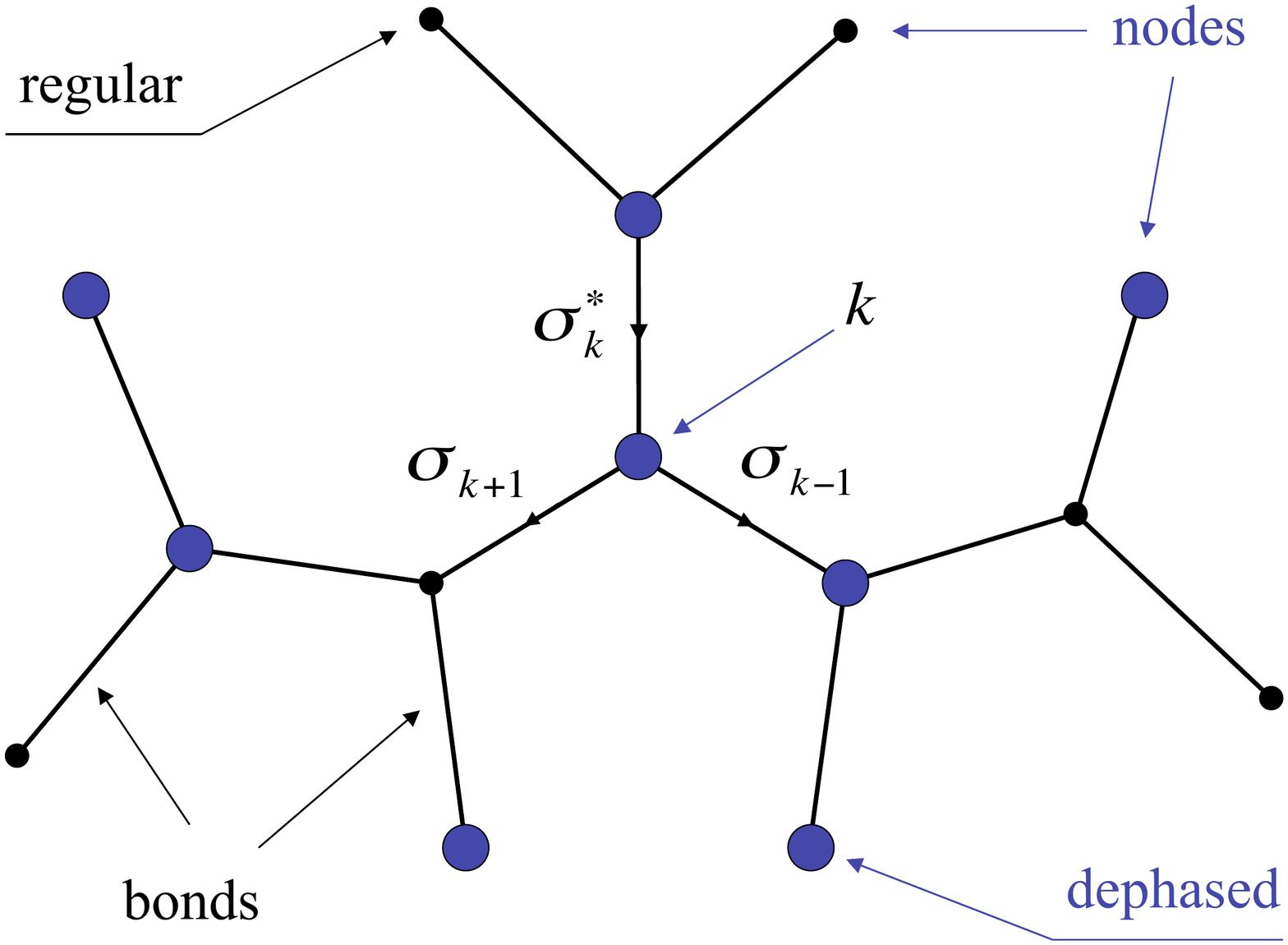}
\caption{\label{} Mapping Eqs.~(\ref{9}) on a Cayley tree. Each node represents a nonlinear eigenstate, or nonlinear oscillator with the equation of motion $i\dot{\sigma}_k - \omega_k \sigma_k - \beta V_{k, k, k, k} \sigma_{k} \sigma^*_{k} \sigma_{k} = 0$. Blue nodes represent oscillators in a chaotic (``dephased") state. Black nodes represent oscillators in regular state. One ingoing and two outgoing bonds on node $k$ ($k= 1,2,\dots$) represent respectively the complex amplitudes $\sigma^*_{k}$, $\sigma_{k-1}$, and $\sigma_{k+1}$.}
\end{figure}

A Cayley tree being by its definition a hierarchical graph (e.g., Ref. \cite{Schroeder}) offers a suitable geometric model for infinite-dimensional spaces. We think of this graph as embedded in Hilbert space, characterized by its metric in Eq.~(\ref{Inn}). 
In the thermodynamic limit $k_{\max}\rightarrow\infty$, in place of a Cayley tree, one uses the notion of a Bethe lattice instead: Its nodes host nonlinear oscillators defined by Eq.~(\ref{eq}); its bonds, conduct oscillatory processes to their nearest neighbors as a result of the interactions present. We assume that each oscillator can be in a chaotic (``dephased") state with the probability $p$ (hence, in a regular state with the probability $1-p$). The $p$ value being smaller than 1 implies that the domains of random motions occupy only a fraction of the lattice nodes. Whether an oscillator is dephased is decided by Chirikov's resonance-overlap condition; which may or may not be matched on node $k$. We believe that in systems with many coupled degrees of freedom each such ``decision" is essentially a matter of the probability. The choice is random. Focusing on the $p$ value, we consider system-average nonlinear frequency shift 
\begin{equation}
\Delta\omega_{\rm NL} = \beta\langle|\psi_n|^2\rangle_{\Delta n}
\label{Shift} 
\end{equation}
as an effective ``temperature" of nonlinear interaction. It is this ``temperature" that rules over the excitation of the various resonant ``levels" in the system. With this interpretation in mind, we write $p$ as the Boltzmann factor 
\begin{equation}
p = \exp (-\delta\omega / \Delta\omega_{\rm NL}),
\label{B} 
\end{equation}
where $\delta\omega$ is the characteristic energy gap between the resonances. Expanding $\psi_n$ over the basis of linearly localized modes, we have 
\begin{equation}
\langle|\psi_n|^2\rangle_{\Delta n} = \frac{1}{\Delta n} \sum_n \sum_{m_1,m_2} \phi^*_{n,m_1}\phi_{n,m_2} \sigma^*_{m_1} \sigma_{m_2}.
\label{Mean} 
\end{equation}  
The summation here is performed with the use of orthogonality of the basis modes. Combining with Eq.~(\ref{Shift}),
\begin{equation}
\Delta\omega_{\rm NL} =  \frac{\beta}{\Delta n}\sum_m \sigma^*_{m} \sigma_{m}.
\label{Upon} 
\end{equation}
The sum over $m$ is easily seen to be equal to 1 due to the conservation of the probability. Thus, $\Delta\omega_{\rm NL} = \beta / \Delta n$. If the field is spread over $\Delta n$ states, then the distance between the resonant frequencies behaves as $\delta \omega \sim 1/\Delta n$. We normalize units in Eq.~(\ref{1}) to have $\delta \omega = 1/\Delta n$ exactly. One sees that  
\begin{equation}
p = \exp (-1/\beta).
\label{Non-p} 
\end{equation}
This result shows that behavior is {\it non-perturbative} in the pseudochaotic regime. For the vanishing $\beta\rightarrow 0$, the Boltzmann factor $p\rightarrow 0$, implying that all oscillators are in regular state. In the opposite regime of $\beta\rightarrow+\infty$, $p\rightarrow 1$. That means that all oscillators are dephased and that the random motions span the entire lattice. 

There exists a critical concentration, $p_c$, of dephased oscillators permitting an escape path to infinity for the first time. This critical concentration is nothing else than the percolation transition threshold on a Cayley tree. As the Cayley tree does {\it not} contain loops \cite{Schroeder}, the value of $p_c$ is expressible in terms of the coordination number only \cite{Havlin}, leading to $p_c = 1/(z-1)$. This is an exact result. For $z=3$, the percolation point is at $p_c = 1/2$. We associate the critical value $p_c = 1/2$ with the onset of transport in the DANSE model, Eq.~(\ref{1}). When translated into the $\beta$ values the threshold condition reads 
\begin{equation}
\beta_c = 1/\ln (z-1).
\label{Betac} 
\end{equation}
Setting $z=3$, we have $\beta_c = 1/\ln 2 \approx 1.4427$. This value defines the critical strength of nonlinearity that destroys Anderson localization. For the $\beta$ values smaller than this, the localization persists, despite that the problem is nonlinear. When $\beta \geq 1/\ln 2$, the localization is lost, and the wave field spreads to infinity. 

Our conclusion so far is that the loss of localization is a threshold phenomenon, which requires the strength of nonlinearity be above a certain level. In this respect, the nonlinearity parameter $\beta$ acquires the role of the delocalization control parameter. The onset of unlimited spreading is at $\beta_c = 1/\ln 2$. This value is characteristic of the DANSE model~(\ref{1}) with quadratic nonlinearity. 

\subsubsection{Random walk model} We now turn to predict the second moments for the onset spreading. This task is essentially simplified if one visualizes the transport as a random walk over a system of dephased oscillators. For $p\rightarrow p_c$, this system is self-similar, i.e., fractal. That means that dephased oscillators form arbitrarily large clusters, each presenting the same fractal geometry of the infinite percolation cluster \cite{Havlin}. It is the infinite cluster that conducts unlimited spreading of the wave function on a Bethe lattice. We should stress that the fractal geometry of the clusters is a consequence of the probabilistic character of dephasing.

In random walks on percolation systems one writes the mean-square displacement from the origin as \cite{Havlin,Gefen}
\begin{equation}
\langle (\Delta n) ^2 (t) \rangle \propto t^{2 / (2+\theta)}, \ \ \ t\rightarrow+\infty.
\label{MS} 
\end{equation}
$\theta$ is the index of anomalous diffusion, or the connectivity index, and accounts for the deviation from the usual Fickian diffusion in fractal geometry. In a basic theory of percolation \cite{Stauffer} it is shown that $\theta \geq 0$ for $p\rightarrow p_c$. That is, the mean-square displacement in Eq.~(\ref{MS}) grows slower-than-linear with time. This slowing down of the transport occurs as a result of long-time trappings and delays of the diffusing particles in multiple tipping points and dead-ends of the fractal. Thus, the diffusion is {\it anomalous}, at contrast with homogeneous spaces; where by definition $\theta\equiv 0$. Another common way of writing Eq.~(\ref{MS}) is given by (e.g., Ref. \cite{Havlin}) 
\begin{equation}
\langle (\Delta n) ^2 (t) \rangle \propto t^{d_s / d_f}, \ \ \ t\rightarrow+\infty,
\label{MS+} 
\end{equation}
where $d_f$ is the Hausdorff dimension, which measures the number of nodes that belong to a given cluster, and $d_s = 2d_f / (2+\theta)$ is the fracton, or spectral, dimension, which describes the density of states in fractal geometry \cite{AO,Rammal}. It also appears in the probability of the random walker to return to the origin ($\propto t^{-d_s / 2}$) \cite{Bouchaud,Shaugh}. The key difference between the Hausdorff and the spectral dimensions lies in the fact that $d_f$ is a purely structural characteristic of the fractal; whereas $d_s$ involves via $\theta$ the dynamical properties, such as wave excitation, diffusion, etc. Note that, because $\theta \geq 0$ for threshold percolation, the spectral dimension is not larger than its Hausdorff counterpart, i.e., $d_s \leq d_f$. The value of $d_s$ can conveniently be considered as the effective fractional number of the degrees of freedom in fractal geometry \cite{Chapter,Bouchaud}, since it naturally substitutes the embedding (integer) dimensionality in respective diffusion \cite{Gefen,Shaugh} and wave-propagation \cite{AO,Rammal,Naka} problems on fractals. 

The two scaling laws above, Eqs.~(\ref{MS}) and~(\ref{MS+}), apply to any percolation system. For percolation on a Cayley tree, also recognized as the mean-field percolation problem \cite{Schroeder,Naka}, the following exact results hold \cite{Havlin,Coniglio}: $\theta =4$, $d_f = 4$, and $d_s = 4/3$. One sees that 
\begin{equation}
\langle (\Delta n) ^2 (t) \rangle \propto t^{1/3}, \ \ \ t\rightarrow+\infty.
\label{MS++} 
\end{equation}
This is the desired scaling. By its derivation, subdiffusion in Eq.~(\ref{MS++}) is asymptotic in the thermodynamic limit. 

We proceed with a remark that the Hausdorff dimension being equal to $d_f = 4$ matches with the implication of Eq.~(\ref{5}) where the coefficients $V_{k, m_1, m_2, m_3}$ are supposed to run over 4-dimensional subsets of the ambient Hilbert space. Indeed it is the overlap integral of four Anderson eigenmodes in Eq.~(\ref{5}), that decides the fractal dimensionality of subsets of phase space on which the transport of the wave field occurs. When the nearest-neighbor rule is applied, this overlap structure is singled out for dynamics. Under the condition that the structure is critical, i.e., ``at the edge" of permitting a path to infinity, the fractal support for the transport is reduced to a percolation cluster on the Bethe lattice. The latter is characterized, along with the above value of the Hausdorff fractal dimension, by the very specific connectivity index, $\theta = 4$. The end result is $\alpha = 2 / (2+\theta)  = 1/3$. 

\subsubsection{Non-Markovian diffusion equation} At contrast with Eq.~(\ref{7}), random walks on percolation systems are described by a non-Markovian diffusion equation with a powerlaw memory kernel \cite{Chapter,PRE09,EPL} 
\begin{equation}
\frac{\partial}{\partial t} f (t, \Delta n) = \frac{1}{\Gamma (\alpha)} \frac{\partial}{\partial t}\int _{0}^{t} \frac{W_\alpha\, dt^{\prime}}{(t - t^{\prime})^{1-\alpha}} \frac{\partial^2}{\partial (\Delta n)^2} f (t^\prime, \Delta n),\label{23} 
\end{equation}
where $\alpha$ ($0 < \alpha < 1$) determines the exponent of the powerlaw (hence the memory decay rate); 
$W_\alpha$ collects parameters of the transport process; and we have chosen $t=0$ as the beginning of the system's time evolution. We associate the Laplace convolution in Eq.~(\ref{23}) with long-time correlated dynamics near separatrix. The differintegration on the right-hand side has the analytical structure \cite{Podlubny,Uch} of fractional time the so-called Riemann-Liouville fractional derivative, ${\partial^{1-\alpha}}/{\partial t^{1-\alpha}}$, 
\begin{equation}
W_\alpha^{-1}\frac{\partial}{\partial t} f (t, \Delta n) = \frac{\partial^{1-\alpha}}{\partial t^{1-\alpha}} \frac{\partial^2}{\partial (\Delta n)^2} f (t, \Delta n).\label{24} 
\end{equation}
In writing diffusion Eq.~(\ref{23}) we have adopted results of Refs.~\cite{PRE09,NJP} to random walks on a single cluster. The fractional order of time differentiation in Eq.~(\ref{24}) is determined by the connectivity index through $1-\alpha = \theta / (2+\theta)$ and is exactly zero for $\theta = 0$. Then the fractional Riemann-Liouville derivative of zero order is unity operator, implying that no fractional properties come into play for homogeneous spaces. Equation~(\ref{23}) when account is taken for the initial value problem can be rephrased \cite{PRE09} in terms of the Caputo fractional derivative \cite{Podlubny,Uch} which shows a better behavior under transformations. Taking moments of the fractional diffusion equation~(\ref{24}) leads to the dispersion law in Eq.~(\ref{MS}), with $\alpha = 2 / (2+\theta)$. Setting $\theta = 4$, one also sees that the critical spreading requires fractional Riemann-Liouville operator of order $\theta / (2+\theta) = 2/3$ for $\alpha = 1/3$. This behavior is asymptotic. The net result is that the diffusion process at criticality is essentially non-Markovian with powerlaw correlations. 

The non-Markovianity is introduced geometrically via the complexity features (contained in the non-zero $\theta$ value) of the infinite percolation cluster. Equation~(\ref{24}) shows that the critical spreading is a matter of {fractional} (or ``strange") kinetics \cite{Klafter,Nature}, consistently with the implication of pseudochaotic behavior \cite{Chapter,Report,JMPB}. Indeed equations built on fractional derivatives offer an elegant and powerful tool to describe anomalous transport in complex systems \cite{Report,Klafter}. There is an insightful connection with a generalized master equation formalism along with a mathematically convenient way for calculating transport moments as well as solving initial and boundary value problems \cite{Klafter,Rest}. 

The fundamental solution of the fractional Eq.~(\ref{24}) is evidenced in Table~1 of Ref.~\cite{Rest}. It shares non-Gaussianity with Green's function in Eq.~(\ref{8}), being in the rest analytically very different.  

A statistical analysis of subdiffusion in the nonlinear Schr\"odinger equation with disorder by means of fractional diffusion equation~(\ref{24}) has been also suggested in Ref. \cite{Iomin}. The idea was that nonlinearity-induced overlap between components of the wave field introduces a distribution of waiting times to the hopping motion, making it possible to utilize the scheme of continuous time random walks \cite{Klafter,Ctrw1}. In the above next-neighbor picture of the transport we have {\it not} as a matter of fact assumed any heavy-tailed distribution of this sort. Indeed, in our model, the random walker is supposed to take one unit step along the cluster as soon as one unit time is elapsed. Even so, with the recognition that dead-ends and other complexity elements of the fractal act as to delay the diffusing particle at all scales (e.g., Refs. \cite{Havlin,Naka}), the critical spreading of the wave field when modeled on a regular lattice can be thought of as corresponding with the effective waiting time distribution, $\chi_{\rm eff} (\Delta t)$, 
\begin{equation}
\chi_{\rm eff} (\Delta t) \propto (1+\Delta t)^{-(4+\theta)/(2+\theta)} \propto (1+\Delta t)^{-4/3},
\label{wtd} 
\end{equation}
thus sustaining the Riemann-Liouville derivative in the fractional diffusion equation~(\ref{24}). Equation~(\ref{wtd}) translates the spatial complexity properties of the fractal structure at percolation into the corresponding waiting-time statistical properties. It is noticed that the fractional diffusion equation in Eq.~(\ref{24}) is ``born" within the exact mathematical framework of nonlinear Schr\"odinger equation with usual time differentiation. Indeed, no {\it ad hoc} introduction of fractional time differentiation in the dynamic Eq.~(\ref{1}) has been assumed to obtain this sub\-diffusion. It is, in fact, the interplay between nonlinearity and randomness, which leads to a non-Markovian transport of the wave function at criticality, and to a time-fractional kinetic equation in the end. 

\section{Arbitrary power nonlinearity} The considerations above can be extended, so that they include a generalized DANSE model with arbitrary power nonlinearity, i.e.,
\begin{equation}
i\hbar\frac{\partial\psi_n}{\partial t} = \hat{H}_L\psi_n + \beta |\psi_n|^{2s} \psi_n,
\label{1s} 
\end{equation}
where $s$ ($s > 0$) is a real number. We define the power $2s$ of the modulus of the wave field as the power $s$ of the probability density, i.e., 
\begin{equation}
|\psi_n|^{2s} \equiv \left[\psi_n\psi_n^*\right]^s.
\label{PNL} 
\end{equation}
Then in the basis of linearly localized modes we can write, with the use of $\psi_n = \sum_m \sigma_m \phi_{n,m}$, 
\begin{equation}
|\psi_n|^{2s} = \left[\sum_{m_1,m_2} \sigma_{m_1} \sigma^*_{m_2} \phi_{n,m_1}\phi^*_{n,m_2}\right]^s.
\label{2s} 
\end{equation}
It is convenient to consider the expression on the right-hand side as a functional map
\begin{equation}
\hat \mathrm{F}_s:\{\phi_{n,m}\}\rightarrow \left[\sum_{m_1,m_2} \sigma_{m_1} \sigma^*_{m_2} \phi_{n,m_1}\phi^*_{n,m_2}\right]^s
\label{3s} 
\end{equation}
from the vector field $\{\phi_{n,m}\}$ into the scalar field $|\psi_n|^{2s}$. It is noticed that the map in Eq.~(\ref{3s}) is positive definite, and that it contains a self-similarity character in it, such that by stretching the basis vectors (by a stretch factor $\lambda$) the value of $\hat \mathrm{F}_s$ is just renormalized (multiplied by $|\lambda|^{2s}$). We have, accordingly,  
\begin{equation}
\hat \mathrm{F}_s\{\lambda \phi_{n,m}\} = |\lambda|^{2s} \hat \mathrm{F}_s\{\phi_{n,m}\}.
\label{4s} 
\end{equation}
Consider expanding the powerlaw on the right-hand side of Eq.~(\ref{2s}). If $s$ is a positive integer, then a regular expansion can be obtained as a sum over $s$ pairs of indices $(m_{1,1}, m_{1,2})\dots (m_{s,1}, m_{s,2})$. The result is a homogeneous polynomial, an $s$-quadratic form \cite{Cox}. In contrast, for fractional $s$, a simple procedure does not exist. Even so, with the aid of Eq.~(\ref{4s}), one might circumvent the problem by proposing that the expansion goes as a homogeneous polynomial whose nonzero terms all have the same degree $2s$. 
``Homogeneous" means that every term in the series is in some sense representative of the whole. Then one does not really need to obtain a complete expansion of $\hat \mathrm{F}_s$ in order to predict dynamical laws for the transport, since it will be sufficient to consider a certain collection of terms which by themselves completely characterize the algebraic structure of $\hat \mathrm{F}_s$ as a consequence of the homogeneity property. 
We dub this collection of terms the backbone, and we define it through the homogeneous map 
\begin{equation}
\hat \mathrm{F}^\prime_s:\{\phi_{n,m}\}\rightarrow \sum_{m_1,m_2} \sigma_{m_1}^s \sigma^{*s}_{m_2} \phi^s_{n,m_1}\phi^{*s}_{n,m_2}.
\label{3ss} 
\end{equation}
In what follows, we consider the backbone as representing the algebraic structure of $\hat \mathrm{F}_s$ in the sense of Eq.~(\ref{4s}). So, for fractional $s$, our analyses will be based on a reduced model which is obtained by replacing the original map $\hat \mathrm{F}_s$ by the backbone map $\hat \mathrm{F}^\prime_s$. The claim is that the reduction $\hat \mathrm{F}_s\rightarrow\hat \mathrm{F}^\prime_s$ does not really alter the scaling exponents behind the wave-spreading, since the algebraic structure of the original map is there anyway. Note that $\hat \mathrm{F}_s$ and $\hat \mathrm{F}^\prime_s$ both have the same degree $2s$, which is the sum of the exponents of the variables that appear in their terms. Note, also, that the original map coincides with its backbone in the limit $s\rightarrow 1$. This property illustrates the significance of the quadratic nonlinearity {\it vs.} arbitrary power nonlinearity.  
Turning to DANSE~(\ref{1s}), if we now substitute the original power nonlinearity with the backbone map, in the orthogonal basis of the Anderson eigenstates we find, with $\hbar = 1$, 
\begin{equation}
i\dot{\sigma}_k - \omega_k \sigma_k = \beta \sum_{m_1, m_2, m_3} V_{k, m_1, m_2, m_3} \sigma_{m_1}^s \sigma^{*s}_{m_2} \sigma_{m_3},
\label{4s+} 
\end{equation}
where 
\begin{equation}
V_{k, m_1, m_2, m_3} = \sum_{n} \phi^*_{n,k}\phi_{n,m_1}^s\phi^{*s}_{n,m_2}\phi_{n,m_3}
\label{5s+} 
\end{equation}
are complex coefficients characterizing the overlap structure of the nonlinear field, and we have reintroduced the eigenvalues of the linear problem, $\omega_k$, satisfying $\hat H_L \phi_{n,k} = \omega_k \phi_{n,k}$. Although obvious, it should be emphasized that the use of the backbone map $\hat \mathrm{F}^\prime_s$ in place of the original map $\hat \mathrm{F}_s$ preserves the Hamiltonian character of the dynamics, but with a different interaction Hamiltonian, $\hat H_{\rm int}$,    
\begin{equation}
\hat H_{\rm int} = \frac{\beta}{1+s} \sum_{k, m_1, m_2, m_3} V_{k, m_1, m_2, m_3} \sigma^*_k \sigma_{m_1}^s \sigma^{*s}_{m_2} \sigma_{m_3}
\label{6s+} 
\end{equation}
extending its quadratic counterpart in Eq.~(\ref{6+}) to $s\ne 1$. Note that $\hat H_{\rm int}$ includes self-ineractions through the diagonal elements $V_{k, k, k, k}$. Another important point worth noting is that the strength of the interaction vanishes in the limit $s\rightarrow\infty$ (as $\sim 1/s$). Therefore, keeping the $\beta$ parameter finite, and letting $s\rightarrow \infty$, one generates a regime where the nonlinear field is asymptotically localized. One sees that high-power nonlinearities act as to reinstall the Anderson localization. We shall confirm this by the direct calculation of respective transport exponents. Equations~(\ref{4s+}) define a system of coupled nonlinear oscillators with a parametric dependence on $s$. Similarly to the DANSE model with quadratic power nonlinearity, each nonlinear oscillator with the Hamiltonian   
\begin{equation}
\hat h_{k} = \omega_k \sigma^*_k \sigma_k + \frac{\beta}{1+s} V_{k, k, k, k} \sigma^*_k \sigma_{k}^s \sigma^{*s}_{k} \sigma_{k}
\label{6+h+s} 
\end{equation}
and the equation of motion 
\begin{equation}
i\dot{\sigma}_k - \omega_k \sigma_k - \beta V_{k, k, k, k} \sigma^s_{k} \sigma^{*s}_{k} \sigma_{k} = 0
\label{eq+s} 
\end{equation}
represents one nonlinear eigenstate in the system $-$ identified by its wave number $k$, unperturbed frequency $\omega_k$, and nonlinear frequency shift $\Delta \omega_{k} = \beta V_{k, k, k, k} \sigma^s_{k} \sigma^{*s}_{k}$. We reiterate that non-diagonal elements $V_{k, m_1, m_2, m_3}$ characterize couplings between each four eigenstates with wave numbers $k$, $m_1$, $m_2$, and $m_3$. The comprehension of Hamiltonian character of the dynamics paves the way for a consistency analysis of the various transport scenarios behind the Anderson localization problem (with the topology of resonance overlap taken into account) \cite{EPL}. To this end, the transport problem for the wave function becomes essentially a topological problem in phase space. With these implications in mind, we consider separately the chaotic and the pseudochaotic cases, regarded as the essential key elements to the dynamics. 

\begin{figure}
\includegraphics[width=0.51\textwidth]{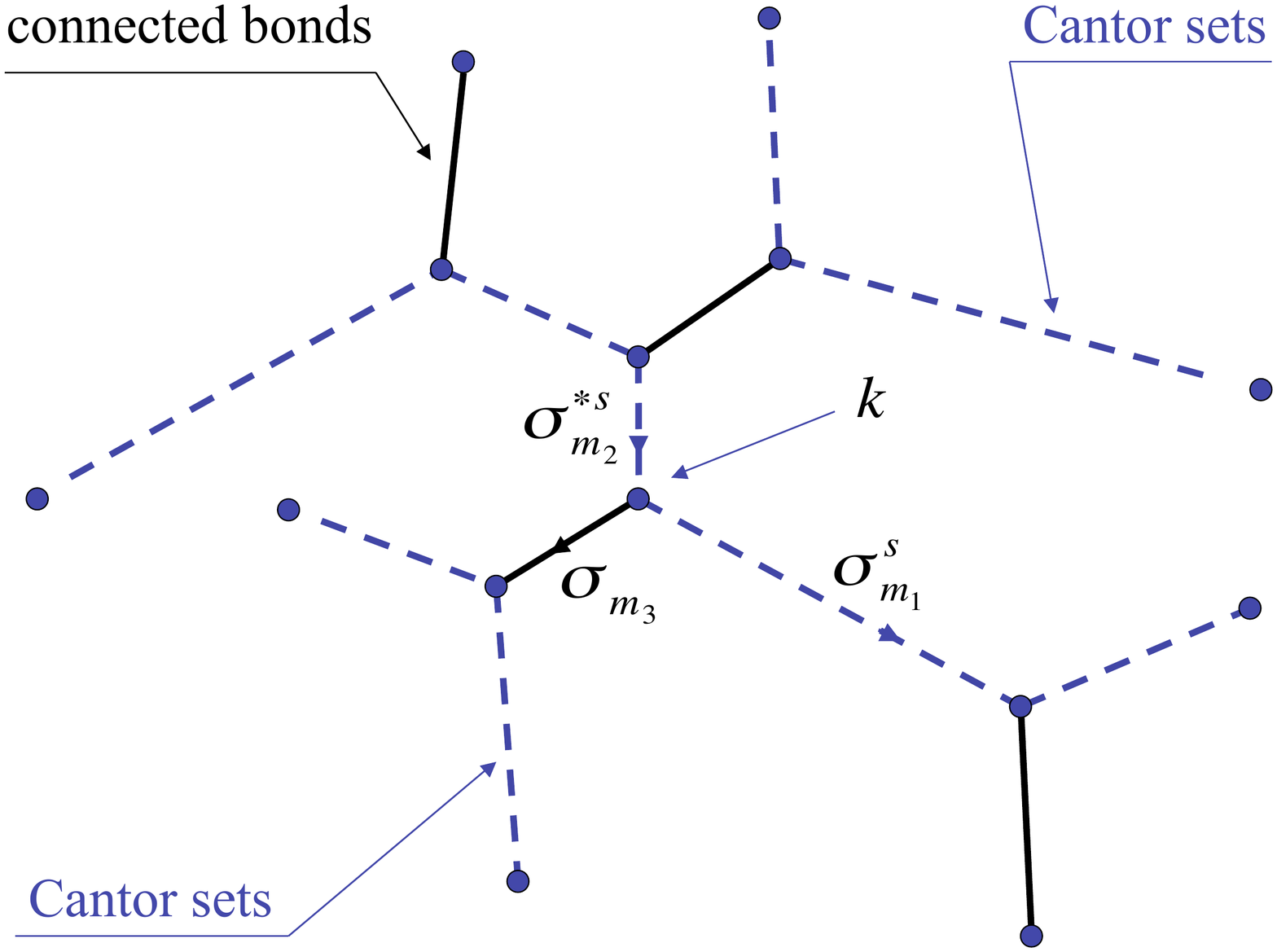}
\caption{\label{} Mapping Eqs.~(\ref{4s+}) on a graph in wave number space. Each node with the coordinate $k$ represents a nonlinear oscillator with the equation of motion $i\dot{\sigma}_k - \omega_k \sigma_k - \beta V_{k, k, k, k} \sigma^s_{k} \sigma^{*s}_{k} \sigma_{k} = 0$. The terms raised to the power $s$, with $0 < s < 1$, generate disconnected bonds. These are Cantor sets with the Hausdorff dimensionality $s$. The disconnected bonds are plotted as dashed lines, also distinguished by their blue color. There are exactly three bonds at each node: one connected bond, which corresponds to the amplitude $\sigma_{m_3}$; and two disconnected bonds, corresponding to respectively the amplitudes $\sigma_{m_1}^s$ and $\sigma^{*s}_{m_2}$. One sees that the average concentration of the connected bonds is equal to $1/3$. This is smaller that the percolation threshold, $p_c = 1/2$, on a Cayley tree with the coordination number $z=3$. Thus, there is no a connected structure, already from the outset, to host a transport of the wave field. As a result, the nonlinear field is Anderson localized similarly to the linear case. Note that we do not assume next-neighbor couplings here, so that lengths of the bonds vary.}
\end{figure}

\subsection{Absence of transport for subquadratic nonlinearity} Focusing on the $s$-dependence of the nonlinear term, we note that only values corresponding to the quadratic and superquadratic nonlinearities, i.e., $s\geq 1$, can cause the wave field to spread. So, there is a threshold power nonlinearity, the quadratic dependence with $s=1$, to enable transport of the wave field. 
No spreading process is to be expected for $s$ a fraction between 0 and 1. This follows from the topology of the nonlinear terms in Eq.~(\ref{4s+}). In fact, when represented geometrically by drawing a graph in wave number space, the terms raised to the power $s$ will correspond to Cantor sets. Their dimensionality (Hausdorff or box-counting) is fractional, and is given by the value of $s$ ($0 < s < 1$). A bond being a Cantor set implies that it is disconnected, so that it cannot transmit interactions. There will be exactly two such bonds at each node: generated respectively by the terms $\sigma_{m_1}^s$ and $\sigma^{*s}_{m_2}$. The remaining bond, the one introduced by the term $\sigma_{m_3}$, is connected, and is available for the dynamics. Even so, it will be the only connected bond here, as dictated by the topology of the nonlinearity (see Fig.~2), hence it cannot transmit the interactions further. The average concentration of the connected bonds appears in the proportion $1:3$ (one connected bond out of three at every node). One sees that it goes below the bond percolation threshold, $p_c = 1/2$, for a Cayley tree with the coordination number $z=3$. The latter is given by $p_c = 1/(z-1)$ for both site and bond percolation problems \cite{Schroeder}. Thus, there is no a connected structure, already from the outset, to host a transport of the wave field. 

The net result is that the field is Anderson localized, similarly to the linear case. This localized state will survive any strength of nonlinearity, contained in the parameter $\beta$. 

Another way to obtain this result is to notice that, with the disconnected bonds left out of the lattice, the interaction problem defined by Eq.~(\ref{4s+}) is effectively a linear problem in that it would involve only one term on the right-hand side, $\sigma_{m_3}$, which is not coupled with any other term to a chain. Then the main impact the interactions will have onto the dynamics is to generate a shift to the on-site energies $\varepsilon_n$, not to drive a field-spreading. All in all, a power nonlinearity with $0 < s < 1$ proves to be too weak to make it with randomness, so that the phenomena of Anderson localization will be characteristic of the nonlinear field, regardless of the $\beta$ value. As a consequence, there is no a localization-delocalization transition here, nor any signature of critical behavior. 

In subsections B$-$D below we shall assume that the $s$ value is at least not smaller than 1, i.e., $s \geq 1$. We shall revise this assumption in subsection E, where a DANSE model with {modified} nonlinearity is considered.      

\subsection{Chaotic case} The main idea here is to consider Eqs.~(\ref{4s+}) as Langevin equations, with the nonlinear term on the right-hand side thought as a driving noise term (Gaussian or other). 

\subsubsection{Gaussian noise term and the diffusion limit} If the couplings are random, an assumption readily justifiable \cite{Flach} for the Anderson problem, and they correspond to situations applicable for the central limit theorem, then the noise term can be taken as a Gaussian white noise in the limit $t\rightarrow+\infty$. As a consequence, the behavior is of the diffusion type. The scaling of the diffusion coefficient is obtained as the intensity of the noise, yielding 
\begin{equation}
D_s \propto \left|\sum_{m_1, m_2, m_3} V_{k, m_1, m_2, m_3} \sigma_{m_1}^s \sigma^{*s}_{m_2} \sigma_{m_3}\right|^2.
\label{Diff} 
\end{equation}
One sees that $D_s \propto |\sigma_n|^{2(2s + 1)}$. Using here that the conservation of the probability requires $|\sigma_n| ^2 \sim 1 / \Delta n$, we have $D_s \propto 1/(\Delta n)^{2s + 1}$, leading to a range-dependent diffusion equation [cf. Eq.~(\ref{7})]
\begin{equation}
\frac{\partial}{\partial t} f (t, \Delta n) = \frac{\partial}{\partial \Delta n}\left[W_s \frac{1}{(\Delta n)^{2s+1}}\frac{\partial}{\partial \Delta n} f (t, \Delta n)\right],
\label{7s} 
\end{equation}
where all dimensional parameters have been absorbed into the definition of $W_s$. The fundamental solution or Green's function of Eq.~(\ref{7s}) is given by
\begin{equation}
f (t, \Delta n) = \frac{(2s+3)/\Gamma \left[1/(2s+3)\right]}{\left[W_s(2s+3)^2t\right]^{1/(2s+3)}} \exp\left[-\frac{(\Delta n)^{2s+3}}{W_s(2s+3)^2t}\right],
\label{8s} 
\end{equation}
from which a powerlaw growth of the second moments 
\begin{equation}
\langle (\Delta n) ^2 (t) \rangle = \frac{\Gamma \left[3/(2s+3)\right]}{\Gamma \left[1/(2s+3)\right]} \left[W_s(2s+3)^2t\right]^{2/(2s+3)},
\label{8s+} 
\end{equation}
can be deduced for $t\rightarrow+\infty$. In the above we have used the natural normalization $\int_0^\infty f (t, \Delta n) d\Delta n = 1$. The end result is that $\langle (\Delta n) ^2 (t) \rangle \propto t^{2/(2s+3)}$ generalizing the $2/5$ behavior to $s > 1$. When $s\rightarrow\infty$, the transport exponent vanishes (as $\sim 1/s$), consistently with the vanishing of $\hat H_{\rm int}$. So, the asymptotic ($s\rightarrow\infty$) behavior corresponds with the Anderson localization taking place, as expected. 

\subsubsection{Going with the L\'evy flights: Local transport revisited} We should stress that the diffusive picture of the wave-spreading is based on the assumption of Gaussianity of the driving noise process, which, in its turn, relies on the central limit theorem. Indeed the ``central limit theorem" of the theory of the probability states that, under certain rather weak conditions, the probability distribution of a sum of random variables is Gaussian (or normal) \cite{Van}. Even so, the property of Gaussianity is not at all trivial, and it does not always hold true; one example of this is a setting where the couplings are random, but the distribution of coupling strengths is devoid of second and higher moments, a situation applicable for the generalized central limit theorem and L\'evy (stable) processes \cite{Gnedenko,Levy}. 

A L\'evy statistical case can occur naturally via back-reaction of the wave process on the oscillatory medium posing long-range spatial correlations through dynamics \cite{Chapter,UFN}. At this point, the assumption that the nonlinear term in Eqs.~(\ref{4s+}) is taken as a Gaussian white noise can be relaxed. Then a suitable model here will consider Eqs.~(\ref{4s+}) as Langevin equations with the driving noise term of the L\'evy type. We take this to be a white L\'evy noise with L\'evy index $2\mu$ ($0 < 2\mu\leq 2$). By white L\'evy noise we mean a stationary random process, such that the corresponding motion process, i.e., the time integral of the noise, is a symmetric $\mu$-stable L\'evy process with stationary independent increments and stretched Gaussian (or stretched exponential, for $0 < 2\mu < 1$) characteristic function \cite{Gnedenko,Levy}. As is well known, the motion process satisfying these criteria is characterized by a broad distribution of jump lengths and is conventionally referred to as ``L\'evy flights" \cite{Klafter,Chechkin,Rest}. Note that the introduction of L\'evy flights in wave number space does not violate physical principles, in contrast to dealing with massive particles in real space, where physics implies a finite velocity of propagation. 

A peculiar feature of L\'evy processes arising in the nonlinear Anderson problem is range-dependence of the noise. This is associated with the nonlinear interaction between the components of the wave field and is manifest in the scaling behavior $D_s \propto 1/(\Delta n)^{2s + 1}$ of the diffusion coefficient. In L\'evy statistics, one accommodates the nonhomogeneous transport by assuming that the terms determining the jump length $|\Delta n - \Delta n^\prime|$ separate from the spatial asymmetry due to $D_s \propto 1/(\Delta n)^{2s + 1}$, implying that the intensity of the L\'evy noise is calculated at the arrival site $\Delta n$ and not at the departure site $\Delta n^\prime$. 
Technically, the separation of terms is implemented based on the generic functional form \cite{Chechkin,Barkai} of the transfer kernel, using the Heaviside step function to ascribe the dependence on the jump length. With these implications in mind, one obtains the following L\'evy-fractional diffusion equation for the probability density, $f = f (t, \Delta n)$:  
\begin{equation}
\frac{\partial}{\partial t} f (t, \Delta n) = \frac{\partial^{2\mu}}{\partial |\Delta n|^{2\mu}}\left[W_{s}^{(\mu)} \frac{1}{(\Delta n)^{2s+1}} f (t, \Delta n)\right].
\label{FD} 
\end{equation}
The symbol $\partial^{2\mu} / \partial |\Delta n|^{2\mu}$ represents fractional differentiation along the coordinate $\Delta n$ and is defined through 
\begin{equation}
\frac{\partial^{2\mu}}{\partial |\Delta n|^{2\mu}} \Phi (t, \Delta n) = \frac{1}{\Gamma_\mu}\frac{\partial^2}{\partial (\Delta n)^2} \int_{-\infty}^{+\infty}\frac{\Phi (t, \Delta n^\prime)}{|\Delta n-\Delta n^\prime|^{2\mu - 1}} d \Delta n^\prime
\label{Def} 
\end{equation} 
for $1 < 2\mu \leq 2$, with $\Gamma_\mu = - 2\cos(\pi\mu)\Gamma(2-2\mu)$; and similarly for $0 < 2\mu < 1$ \cite{Uch,Chechkin}. 
The two-sided improper integral is understood as the sum $\int_{-\infty}^{\Delta n} + \int_{\Delta n}^{+\infty}$. In the above $\Phi (t, \Delta n)$ belongs to the class of differintegrable functions \cite{Klafter,Oldham}, which can be expanded into a power series with an algebraic leading term. One sees that $\partial^{2\mu} / \partial |\Delta n|^{2\mu}$ is an integro-differential operator, which has the analytical structure of ordinary space differentiation acting on a Fourier convolution of the function $\Phi (t, \Delta n)$ with a powerlaw. It interpolates between a pure derivative and a pure integral, and is often referred to as the fractional Riesz operator. By its definition, the Riesz operator can conveniently be considered as a normalized sum of left and right Riemann-Liouville derivatives on the infinite axis \cite{Klafter,Samko}. It is this operator, which incorporates the nonlocal properties of the transport. In the Gaussian limit $2\mu = 2$, the Riesz operator reduces to the conventional Laplacian, so that local behavior is reproduced. Further $D_s \propto 1/(\Delta n)^{2s + 1}$ allows for non-homogeneous transport and absorbs in a single scaling dependence the radial decay of the driving noise term. To this end, fractional diffusion equation~(\ref{FD}) represents a competition between nonlocality (contained in the Riesz derivative and the fractional exponent $0 < 2\mu < 2$) and nonhomogeneity of the transport (contained in the power $2s$). 

L\'evy-fractional equations of the diffusion and Fokker-Planck type including nonhomogeneous ventures have been explored for a large variety of systems and of physical conditions (e.g., Refs. \cite{JJR,Chechkin,Barkai,Fog,Jesper,Gonchar}; reviewed in Refs. \cite{Report,Klafter,Rest,Uch}). 
Note that the fractional diffusion equation~(\ref{FD}) is Markovian, in contrast to Eq.~(\ref{23}) discussed above, and that it involves nonlocal differentiation over the space, rather than the time, variable, consistently with the chaotic property of the dynamics. Mathematically, nonlocal equations with range-dependent diffusion coefficient have been considered in Ref. \cite{Srok}, where one can also find their solutions in terms of the Fox $H$-function. Even so, the basic physics significance of these equations has not been unveiled. Here, we propose that the range-dependence occurs naturally as a consequence of nonlinear interaction between the components of the wave field and is governed by the power nonlinearity generating DANSE. 
This competition between nonlocality and nonhomogeneity is the defining feature of the proposed transport model. The scaling of the second moments is given by
\begin{equation}
\langle (\Delta n) ^2 (t) \rangle \propto t^{2 / (2\mu + 2s + 1)}, \ \ \ t\rightarrow+\infty.
\label{MS-ext} 
\end{equation}
Setting $\mu = 1$, one recovers the local behavior in Eq.~(\ref{8s+}) above. Because of the strong range-dependence, contained in $D_s \propto 1/(\Delta n)^{2s + 1}$, the transport is {subdiffusive} for all $0 < \mu \leq 1$ despite that the dynamics are nonlocal; an apparent challenge to the classical vision of the L\'evy motion as a paradigmatic model \cite{Klafter,Chechkin} of superdiffusion. Thus, nonhomogeneity, posed by the nonlinear interaction, effectively slows down the nonlocal transport. On the other hand, nonlocal behavior acts as to significantly enhance the field-spreading on L\'evy flights when compared to local transport case. We consider these nonlocal transport regimes ``anomalous," just to address a natural analogue here with the issue of anomalous transport in plasmas and fluids \cite{Report,Bouchaud,UFN}. 

There is an upper bound on anomalous spreading in nonlinear Schr\"odinger models, and this corresponds to a setting with $\mu\rightarrow 0$ and $s\rightarrow 1$. The former limit, $\mu\rightarrow 0$, constitutes one important condition on stability of the L\'evy motion \cite{Gnedenko}. The latter limit, $s\rightarrow 1$, merely says that the nonlinearity is quadratic. Then from the general scaling law in Eq.~(\ref{MS-ext}) one obtains 
\begin{equation}
\langle (\Delta n) ^2 (t) \rangle \propto t^{2/3}, \ \ \ t\rightarrow+\infty. 
\label{MS+ext} 
\end{equation}
Equation~(\ref{MS+ext}) presents an upper bound on the rate of nonlocal field-spreading in nonlinear Schr\"odinger models with superquadratic power nonlinearity ($s\geq 1$). 

\subsection{Pseudochaotic case} The ``edge" character of pseudochaotic behavior corresponds to infinite chains of next-neighbor interactions with a minimized number of links at every step. For the reasons of symmetry, when summing on the right-hand side of Eq.~(\ref{4s+}), the only combinations of terms to be taken into account, apart from the self-interaction term $\sigma_{k}^s \sigma^{*s}_{k} \sigma_{k}$, are, essentially, $\sigma_{k-1}^s \sigma^{*s}_{k} \sigma_{k+1}$ and $\sigma_{k+1}^s \sigma^{*s}_{k} \sigma_{k-1}$. These terms will come with respective interaction amplitudes $V_{k, k, k, k}$, $V_{k, k-1, k, k+1}$, and $V_{k, k+1, k, k-1}$, which we shall denote simply by $V_k$, $V_k^-$, and $V_k^+$. Then on the right-hand side (r.h.s.) of Eq.~(\ref{4s+}) we have
\begin{equation}
{\rm r.h.s.} = \beta V_{k} \sigma_{k}^s \sigma^{*s}_{k} \sigma_{k} + \beta \sum_{\pm}V_{k}^\pm \sigma_{k\pm1}^s \sigma^{*s}_{k} \sigma_{k\mp1}.
\label{4s++} 
\end{equation} 
The interaction Hamiltonian in Eq.~(\ref{6s+}) becomes
\begin{equation}
\hat H_{\rm int} = \frac{\beta}{1+s} \sum_{k} \left[V_{k} \sigma^*_k \sigma_{k}^s \sigma^{*s}_{k} \sigma_{k} + \sum_{\pm}V_{k}^\pm \sigma^*_k \sigma_{k\pm1}^s \sigma^{*s}_{k} \sigma_{k\mp1}\right] 
\label{6s++} 
\end{equation}
generalizing its partial (i.e., $s=1$) case in Eq.~(\ref{6++}) to arbitrary real power $s\geq 1$. Assuming that the exponent $s$ is confined between two integer numbers, i.e., $j\leq s < j+1$, in the next-neighbor interaction term we can write
\begin{equation}
\hat H^\prime_{\rm int} = \frac{\beta}{1+s} \sum_{k} \sum_{\pm}V_{k}^\pm \left[\sigma^*_k \sigma_{k\pm1}^j \sigma^{*j}_{k} \sigma_{k\mp1}\right] \sigma_{k\pm1}^{s-j} \sigma^{*s-j}_{k},
\label{6s++p} 
\end{equation}
where the prime symbol indicates that we have extracted the self-interactions. When drawn on a graph in wave-number space, the terms raised to the power $s-j$ will correspond to disconnected bonds, thought as Cantor sets with the fractal dimensionality $0\leq s-j < 1$. Hence, they will not contribute to field-spreading. These terms, therefore, can be cut off from the interaction Hamiltonian, suggesting that only those terms raised to the integer power, $j$, should be considered. We have, accordingly,    
\begin{equation}
\hat H^\prime_{\rm int} \rightarrow \frac{\beta}{1+s} \sum_{k} \sum_{\pm}V_{k}^\pm \sigma^*_k \sigma_{k\pm1}^j \sigma^{*j}_{k} \sigma_{k\mp1}.
\label{6s++c} 
\end{equation}
This is the desired result. Equation~(\ref{6s++c}) defines the effective reduced interaction Hamiltonian in the parameter range of pseudochaotic dynamics for $j\leq s < j+1$. 

Focusing on the transport problem for the wave field, because the interactions are next-neighbor-like, it is convenient to project the system of coupled dynamical equations~(\ref{4s++}) on a Cayley tree, such that each node with the coordinate $k$ represents a nonlinear eigenstate, or nonlinear oscillator with the equation of motion~(\ref{eq+s}); the outgoing bonds represent the complex amplitudes $\sigma_{k\pm 1}$ and $\sigma_{k\mp 1}$; and the ingoing bonds, which involve complex conjugation, represent the complex amplitudes $\sigma_{k}^*$. To make it with the amplitudes $\sigma_{k}^*$ when raised to the algebraic power $s$ one needs for each node a fractional number $s$ of the ingoing bonds. Confining the $s$ value between two nearest integer numbers, $j\leq s < j+1$, we carry on with $j$ connected bonds, which we charge to receive the interactions, and one disconnected bond, which corresponds to a Cantor set with the fractal dimensionality $s-j$, and which cannot transmit the waves. At this point we cut this bond off the tree. A similar procedure applied to the amplitudes $\sigma_{k\pm 1}$, coming up in the algebraic power $s$, generates $j$ outgoing bonds, leaving one disconnected bond behind. Lastly, the remaining amplitude $\sigma_{k\mp 1}$, which does not involve a nonlinear power, contributes with one outgoing bond for each combination of the indexes. One sees that the mapping requires a Cayley tree with the coordination number $z = 2j+1$. The percolation problem on such a tree \cite{Schroeder} leads to a well-defined threshold at $p_c = 1 / (z-1) = 1/2j$. We note in passing that a Cayley tree with $z=3$ constitutes the highest threshold for percolation. Generally, the threshold decays with the order of nonlinearity as $\sim 1/s$, so that the asymptotic behavior, with $s\rightarrow\infty$, is not thresholded.  

\subsubsection{Nonlinear feedback dynamics: Site occupancy decays while spreading
} 
At a first glance, the problem of arbitrary power nonlinearity, with $s > 1$, is similar to the quadratic power case. It is assumed that each nonlinear oscillator, i.e., each node of the Cayley tree, can be in a dephased state with the probability $p$ and in a regular state with the probability $1-p$, and that the distribution of dephased and regular states is essentially random. The actual dynamical state for each oscillator is decided by Chirikov's resonance-overlap conditions \cite{Sagdeev,ChV,Chirikov} whose matching or not matching on a given node $k$ is taken to be a matter of the probability in the limit $k_{\max}\rightarrow\infty$. With these implications in mind, a model for pseudochaotic transport is obtained as a random walk model on connected clusters of dephased oscillators \cite{EPL}. Therefore, it is tempting to look for a localization-delocalization transition as a percolation transition at $p=p_c$, then translate the percolation threshold into a critical value of $\beta$, and finally obtain a subdiffusive spreading of the wave function near criticality.      
%
Even so, this promising program fails at some point, as soon as the parameter $s$ is larger than 1. The main reason for this is that the nonlinear frequency shift, $\Delta \omega_{k} = \beta V_{k, k, k, k} \sigma^s_{k} \sigma^{*s}_{k}$, is not anymore proportional with the distance between the resonances, so that the probability of site occupancy appears to depend on the number of states. Mathematically, to address the statistical significance of the percolation probability, it is convenient to consider the nonlinear frequency shift after averaging over all nodes. We have, with $V_k = V_{k, k, k, k}$, 
\begin{equation}
\Delta\omega_{\rm NL} = \frac{\beta}{\Delta n}\sum_{k=1}^{\Delta n} V_{k} \sigma^s_{k} \sigma^{*s}_{k}, 
\label{sas} 
\end{equation}
leading, when account is taken for the conservation of the probability, to a simple scaling relation 
\begin{equation}
\Delta\omega_{\rm NL} \sim \beta \langle V_k\rangle |\sigma_n|^{2s} \sim \beta \langle V_k\rangle / (\Delta n)^s,
\label{sas+} 
\end{equation}
where angle brackets denote system average. Similarly to the quadratic power case, with $s=1$, we consider $\Delta\omega_{\rm NL}$ as the effective ``temperature" of nonlinear interaction. We note in passing that $\Delta\omega_{\rm NL}$ being inversely proportional with $(\Delta n)^s$ implies that the field is ``cooling off" while spreading, as it should. In this regard, the exponent $s$ characterizes the thermodynamics of the field expansion. Indeed one sees that Eq.~(\ref{sas+}) is nothing else than the thermodynamic equation of state, where $\Delta n$ stands for volume, and $s$ stands for the adiabatic exponent. In a basic thermodynamics of ideal gases \cite{Levich} one writes this exponent as $s = 1+2/\kappa$, where $\kappa$ is the number of degrees of freedom of a molecule, so that for for instance a monoatomic gas, with three degrees of freedom, $\kappa = 3$ and $s= 5/3$. The adiabatic character means that there is no heat and energy exchange with the exterior. In this regard, the interpretation of the exponent $s$ as ``adiabatic" parameter finds a parallel in the two conserved quantities of the DANSE model the Hamiltonian, $\hat H$, and the total probability, $\int |\psi_n|^2 d\Delta n$. One sees that the exponent $s$ being larger than 1 has a strong thermodynamic background, and that the quadratic nonlinearity, identified as $s=1$, is, in fact, very special as it holds only as a limit $\kappa\rightarrow\infty$. When one notices that the characteristic distance between frequencies of excited modes is inversely proportional with the volume, i.e, $\delta\omega \sim 1/\Delta n$, one obtains the probability of site occupancy as the Boltzmann factor, cf. Eq.~(\ref{B}), 
\begin{equation}
p = \exp (-\delta\omega / \Delta\omega_{\rm NL}) = \exp \left[-(\Delta n)^{s-1}/\beta \langle V_k\rangle\right],
\label{B+} 
\end{equation}
where the scaling in Eq.~(\ref{sas+}) has been applied. So it is found for $s\ne 1$ that the Boltzmann factor $p$ depends on the number of occupied states; a remarkable property, which does not find analogues in the quadratic nonlinearity case. For $s>1$, the Boltzmann factor $p$ is a decreasing function of $\Delta n$; that is, $d\Delta n / dt > 0$ implies $dp / dt < 0$. We expect this behavior to have an important impact onto the dynamics. Indeed it is due to the oscillators in dephased state to transmit the wave function between the next-neighbor nodes; so their decaying concentration does not mean other than a progressing deterioration of the transport. In a sense, the spreading process with $s > 1$ has feedback on the dynamical state of the lattice, so that the expansion of the wave field is self-controlling. Another way of putting this is to say that there is an ``aging" through the spreading process. In the thermodynamic limit $\Delta n \rightarrow \infty$, the probability $p$ vanishes, i.e., $p\rightarrow 0$. That means that all oscillators are in regular state, where no transport is possible. Hence, there is a transition to regularity through pseudochaotic feedback dynamics, that blocks spreading beyond a certain limit. 

This transition to regularity occurs exactly at the percolation point, $p=p_c$. Thus, the asymptotic state of the wave field is a quiescent state at the border of regular behavior. No further expansion is allowed based on the next-neighbor rule, reflecting the lack of connectedness in the parameter range $p < p_c$. 

\subsubsection{Field-spreading on finite clusters} Even so, the field can spread to finite domains of wave number space until the $p$ factor reaches its critical value $p_c$ from above. One sees that there is a maximal number of states the field can occupy before the spreading is damped by the $\Delta n$-dependence. This maximal number of states is readily obtained from Eq.~(\ref{B+}) by allowing $p \rightarrow p_c$. So one gets, with $p_c = 1/2j$,  
\begin{equation}
\Delta n_{\max} = \left[-\beta \langle V_k \rangle \ln p_c\right]^{1/(s-1)} = \left[\beta \langle V_k \rangle \ln 2j\right]^{1/(s-1)}.
\label{Bs+} 
\end{equation}
The existence of an upper bound on $\Delta n$ implies that the processes of field-spreading are confined to finite clusters in wave number space. This excludes the possibility of unlimited spreading by pseudochaotic dynamics in the parameter range of superquadratic nonlinearity. If one introduces the entropy, $S = S(p)$, 
\begin{equation}
S (p) = \ln \Delta n (p) = \frac{1}{s-1} \ln \left[-\beta \langle V_k \rangle \ln p\right],
\label{Ent} 
\end{equation}
one obtains for spreading with feedback
\begin{equation}
\frac{d}{dt} S (p) = \frac{1}{s-1} \frac{d}{dt} \ln \ln p = \frac{1}{s-1} \frac{1}{p\ln p} \frac{dp}{dt} \geq 0,
\label{Hth} 
\end{equation}
where use has been made of the general conditions $s > 1$ and $dp/dt \leq 0$. We consider the condition in Eq.~(\ref{Hth}) as the analog Boltzmann's $\mathrm{H}$-theorem for pseudochaotic spreading. Note that the entropy $S (p)$ is maximized at the percolation threshold, yielding 
\begin{equation}
S (p)_{\max} = S(p_c) = \frac{1}{s-1} \ln \left[\beta \langle V_k \rangle \ln 2j\right].
\label{Max} 
\end{equation}
Clearly, the entropy $S (p)_{\max} \rightarrow +\infty$ for $s\rightarrow 1$. We reiterate that the quadratic nonlinearity is a very special case in that it is the only case when the $p$ factor does not involve a dependence on $\Delta n$, permitting unlimited spreading without a feedback. It is understood that the spreading over finite domains corresponds with the second moments growing slower than a powerlaw in the limit $t\rightarrow+\infty$. As no asymptotic spreading occurs for $\Delta n \rightarrow \infty$, transport equations of the Fokker-Planck type are not well-defined here. 

Even so, the typical signatures of anomalous diffusion on fractals and of powerlaw behavior of the moments might be conceived for short times based on random walk approach, using the general dispersion law in Eq.~(\ref{MS}) and an effective connectivity index, $\theta_{\rm eff}$. The power nonlinearity in Eq.~(\ref{1s}) implies that the connectivity value is a {multiplicative} function of $s$. Also one expects this function to naturally reproduce the known mean-field result $\theta = 4$ for threshold percolation on Bethe lattices in the limit $s\rightarrow 1$. Then the obvious dependence satisfying these criteria is $\theta_{\rm eff} = 4s$, where $s\geq 1$. For integer and half-integer $s$, this dependence can alternatively be derived based on combinatorical arguments, using a triangulation procedure in an Euclidean mapping space and the notion of one-bond-connected polyhedron \cite{JSTAT}. So restricting ourselves to times for which the random walker moves only on self-similar scales in the functional Hilbert space, we write, with $\Delta n_{\max} \gg 1$,
\begin{equation}
\langle (\Delta n) ^2 (t) \rangle \propto t^{1 / (2s+1)}, \ \ \ 1\ll t\ll (\Delta n_{\max})^{2(2s+1)},
\label{Finite} 
\end{equation}
from which the scaling dependence in Eq.~(\ref{MS++}) can be deduced for quadratic nonlinearity, yielding $\alpha = 1/3$. One thus predicts that the transition to regularity occurs at time $t_{\rm reg} \sim (\Delta n_{\max})^{2(2s+1)}$ after the initially localized wave packet has been released. Concerning Eq.~(\ref{Finite}), one sees that the power nonlinearity amplifies the complexity properties of the fractal, such as connectivity, dead-ends, etc. \cite{Havlin,Naka}; which act as to slow down the transport on ``anomalous" length scales associated with the self-similar geometry of the clusters. This field-spreading on finite clusters has been already discussed \cite{Iomin,Skokos_PRE} based on computer simulation results, using one-dimensional disordered Klein-Gordon chains with tunable nonlinearity. It is noticed that the exponent of the powerlaw, $\alpha = 1/ (2s+1)$, vanishes in the limit $s\rightarrow \infty$, conformally with the previous considerations.  

The net result of the discussion above consists in the following. When $s\ne 1$, there is a feedback of the spreading process on the phase space of the system. For $s>1$, the operation of this feedback is such, that the concentration of dephased oscillators decays while spreading. Then there is a transition to regular (KAM regime) behavior, which blocks pseudochaotic spreading exactly at the percolation point, $p=p_c$. The asymptotic state of the field is a quiescent state at the border of regularity, characterized by the presence of finite clusters of dephased oscillators. 

Note that $\Delta n = \Delta n (t)$ measures the number of actually excited modes across the wave field and is also related to the initial condition that the wave packet is initially localized. On the contrary, the $p$ factor assesses the likelihood whether a newly involved mode is born in regular or dephased state. In fact, the feedback dynamics dictates that the probability-per-site to be born in a dephased state diminishes with time (for $s>1$). In this regard, the condition for critical behavior in the Hilbert state, i.e., $p\rightarrow p_c$, does {\it not} guarantee that the field has spread to infinity; it merely prescribes a suitable connected pattern of dephased oscillators controlling the rate of the excitation of the field-modes. Also the Boltzmann factor in Eq.~(\ref{B+}) shows that there is {\it always} a critical strength of the nonlinearity parameter, above which the transport is fully chaotic (hence unlimited for $t\rightarrow+\infty$), i.e.,  
\begin{equation}
\beta_c = -\frac{(\Delta n)^{s-1}}{\langle V_k \rangle \ln p_c}.
\label{Ch-P} 
\end{equation}  
Even so, this critical strength is dynamic in general, as it involves a dependence on the number of already occupied states. Hence, a ``universal" critical strength of nonlinear interaction, separating the chaotic and the regular behaviors, only exists for quadratic nonlinearity, with $s=1$; where it is expressible in terms of the percolation threshold according to $p_c = \exp (-1 / \beta_c)$; and is dynamically evolving through the spreading process otherwise as a function of the number of states $\Delta n = \Delta n (t)$. In this regard, the $\beta$ value determines the running 
concentration of dephased oscillators in the Hilbert mapping space as a function of the running $\Delta n (t)$ value; but not really the fact that the field does or does not spread to infinity in wave number space.

\subsubsection{Self-organized criticality in Hilbert space} When considered on a Bethe lattice, a spreading process with $s>1$ is associated with a decaying concentration of dephased oscillators, which, in its turn, is controlled by a dynamically induced nonlinear twist with the number of already occupied states $\Delta n$. An initially homogeneous distribution, which fills the lattice, will change as time progresses in a highly structured, very intermittent distribution near the percolation point, as the lattice order parameter $p$ is progressively lost through the spreading. One sees that the phase space of the DANSE model~(\ref{1s}) is a dynamic medium, which evolves via the next-neighbor readjustments in a marginally connected (i.e., critical percolation) state, beyond which a continued readjustment is not allowed by the lack of connectedness. So, marginal connectedness attracts the nonlinear feedback dynamics. This, together with the fact that the entropy in Eq.~(\ref{Ent}) increases with the decreasing $p$, implies that the operation of feedback {\it automatically} (without tuning of parameters) brings the system in a state of critical percolation on the infinite Cayley tree and that the critical state is robust against variation of parameters. We consider this process as an evolutionary process in the Hilbert space generating via the self-organization an attracting critical point at $p=p_c$. This behavior bears signatures, enabling to associate it with the phenomena of self-organized criticality (SOC) $-$ with that peculiar element that the SOC processes, which we discuss, operate in the functional mapping space, characterized by the inner product in Eq.~(\ref{Inn}). 

The self-organized nature of the criticality stems from the fact that the spreading process with $s\ne 1$ has feedback on the distribution of dephased oscillators in the Hilbert mapping space, providing a back-reaction on the structure on which the transport processes concentrate. We note in passing that the theoretical concept of SOC has been initially applied by Bak {\it et al.} \cite{Bak} to explain fractals and the $1/f$ noise, and has been generalized to nonlinear systems with many coupled degrees of freedom that are driven in a critical state. These topics are summarized in a recent book, Ref. \cite{Ash2013}, and in a review, Ref.  \cite{SSRv}. The analyses, presented here, suggest that the system of dephased oscillators at the percolation onset in the Hilbert mapping space is in SOC state. In this regard, the nonlinear Schr\"odinger models (i.e., the Gross-Pitaevskii equation) with randomness offer a general theoretical framework for mean-field SOC in the parameter range of pseudochaotic behavior. This may turn out to be a very accurate approach to formulate the general analytical theory of SOC, which is underway.
  
Finally, we note that self-organization anomalous phenomena arising from the Gross-Pitaevskii equation with algebraic nonlinearity paves the way to {\it quantum} analogs of SOC; in particular, in the context of superconductivity transition in complex and disordered superconductors \cite{PLA}. The crossover between quantum self-organized criticality and the classical SOC discussed here remains for future investigation.    
   
%
%
\subsection{The front of diffusion: Back to chaotic transport} So far when considering models of pseudochaotic behavior we have relied on the precise picture of next-neighbor random walks on a fractal cluster at percolation. In the practical computer simulations of dynamical field-spreading, however, the limiting conditions behind the threshold percolation might be very difficult to realize, if only due to coarse-graining of parameters defining a system at criticality as well as the natural limitations with respect to finite size effects and the possible lack of the statistics. This can be a cause of further subtlety, which needs to be addressed. 

The subtlety arises, when the random walk problem at percolation is substituted by the front diffusion problem, which is less sensitive to the assumptions of criticality. Going for a model of the front diffusion, we adopt consistently with the mean-field approximation that the structure on which the transport processes concentrate does {\it not} contain loops, so that a minimal-distance the so-called chemical metric coincides with the Pythagorian metric and the front of diffusion is a spherical cut, without contradicting the fact that the Rammal-Toulouse equation holds \cite{Havlin}. Indeed it is found in the basic theory of loopless fractals that the diffusion front propagates according to (see Ref. \cite{Havlin}; p.~200)   
\begin{equation}
\langle (\Delta n) ^2 (t) \rangle \propto t^{2/(d_f + 1)}, \ \ \ t\rightarrow+\infty,
\label{Front} 
\end{equation}
where $d_f$ is the fractal dimension of the chemical path. The overlap integral in Eq.~(\ref{5s+}) suggests that $d_f = 2s+2$. We accept that it is this dimension, characterizing the nonlinear interaction between the components of the wave field, which determines the chemical path for field-spreading. So, using this value in Eq.~(\ref{Front}), one is led to $\langle (\Delta n) ^2 (t) \rangle \propto t^{2/(2s+3)}$, which also recovers the scaling law in Eq.~(\ref{8s+}) above. Setting $s=1$, the ``universal" behavior $\propto t^{2/5}$ is immediately reproduced. The net result is that the front diffusion problem correctly phrases the chaotic regime of field-spreading in the limit $t\rightarrow+\infty$. It does {\it not} rephrase the pseudochaotic scaling $\propto t^{1 / (2s+1)}$ on finite clusters, though. This last observation sheds new light on the different dynamical implications \cite{EPL} behind the $2/5$ and $1/3$ scaling exponents. 

\subsection{Robustness of power nonlinearity: Interaction term modified} Our final note is concerned with {\it robustness} of the DANSE model~(\ref{1s}), where the power nonlinearity, represented by $|\psi_n|^{2s}$, absorbs in a simple dependence the nonlinear interaction between the components of the wave field. Using here that the square of the probability density is a natural order parameter, one finds for $|\psi_n|^2 \rightarrow +0$ 
\begin{equation}
\frac{\partial}{\partial |\psi_n|^2} |\psi_n|^{2s} \propto |\psi_n|^{2(s-1)} \rightarrow +\infty,
\label{Vuln} 
\end{equation}  
implying that the interaction dynamics with $s < 1$ is highly susceptible to small variations in the wave field intensity. On the one hand, this casts doubts on the basic physics significance of the respective Anderson models as opposed to the models with $s\geq 1$. On the other hand, it suggests a simple modification of the DANSE model in Eq.~(\ref{1s}) allowing some transport of the wave field in the parameter range $s < 1$. The main idea here is that the modulus function $|\psi_n|$, which by its definition is neither smooth {nor} analytic, is replaced by $\psi_n + \psi_n^*$, offering via the phase dependence a better behavior under perturbations. Then the modified Anderson model with power nonlinearity is represented by, with $s > 0$, 
\begin{equation}
i\hbar\frac{\partial\psi_n}{\partial t} = \hat{H}_L\psi_n + \beta (\psi_n + \psi_n^*)^{2s} \psi_n.
\label{1s_Mod} 
\end{equation}
This model is remarkable, as it allows for the phenomena of unlimited spreading for all $s \geq 1/2$, and not only for $s \geq 1$, as in the standard model~(\ref{1s}). The demonstration builds upon the fact that the power nonlinearity with $s = 1/2$ naturally accommodates a one-dimensional escape path to infinity via a connected {chain} of dephased oscillators. These chains can always be constructed for $s \geq 1/2$, but not really for $s < 1/2$, so that the borderline regime, with $s = 1/2$, is critical (minimally connected). Expanding the wave function $\psi_n$ over an orthogonal basis of the Anderson eigenstates, and substituting into Eq.~(\ref{1s_Mod}), after a simple algebra one obtains for $s=1/2$ the exact equations of motion $i\dot{\sigma}_k - \omega_k \sigma_k = {\rm r.h.s.}$, $k=1,2,\dots$, where $\omega_k$ are the usual eigenvalues of the linear problem; $\sigma_k$ are the amplitudes of the wave field in the basis of linear localized modes;  
\begin{equation}
{\rm r.h.s.} = \beta \sum_{m_1, m_2} \left[U_{k, m_1, m_2} \sigma_{m_1} \sigma_{m_2} + V_{k, m_1, m_2} \sigma_{m_1} \sigma^*_{m_2}\right];
\label{44} 
\end{equation}
and the complex coefficients $U_{k, m_1, m_2}$ and $V_{k, m_1, m_2}$ are given by, respectively,
\begin{equation}
U_{k, m_1, m_2} = \sum_{n} \phi^*_{n,k}\phi_{n,m_1}\phi_{n,m_2} 
\label{441} 
\end{equation}
and
\begin{equation}
V_{k, m_1, m_2} = \sum_{n} \phi^*_{n,k}\phi_{n,m_1}\phi^*_{n,m_2}.
\label{442} 
\end{equation}
Assuming that the interactions are local (next-neighbor-like), one finds for $s=1/2$ that the onset of unlimited transport corresponds to the one-sided chain reaction 
\begin{equation}
{\rm r.h.s.}^{\prime} = \beta V_{k, k+1, k} \sigma_{k+1} \sigma^*_{k}, \ \ \ k=1,2,\dots,
\label{445} 
\end{equation}
where the complex amplitude $\sigma^*_{k}$ characterizes an ingoing wave process at node $k$, and $\sigma_{k+1}$ characterizes an outgoing process directed to the next node. ``One-sided" means that the chain reaction propagates toward larger wave numbers consistently with the condition that the field is initially localized. 

When appointed geometrically by drawing on a graph, the chain reaction in Eq.~(\ref{445}) corresponds to a ``Cayley tree" with the coordination number $z=2$ and the percolation transition threshold $p_c = 1$. As usual, we consider this tree as embedded in a Hilbert space with metric~(\ref{Inn}). There are exactly {\it two} bonds at each node of the tree: one ingoing bond representing the complex amplitude $\sigma^*_{k}$, and one outgoing bond representing the complex amplitude $\sigma^*_{k}$. Each node represents a nonlinear eigenstate, or nonlinear oscillator with the equation of motion 
\begin{equation}
i\dot{\sigma}_k - \omega_k \sigma_k = U_{k, k, k} \sigma_{k} \sigma_{k} + V_{k, k, k} \sigma_{k} \sigma^*_{k},
\label{447} 
\end{equation}
where $U_{k, k, k}$ and $V_{k, k, k}$ are the diagonal matrix elements defined by their general functional forms in Eqs.~(\ref{441}) and~(\ref{442}). The percolation threshold $p_c$ being equal to 1 implies that the nodes must be absolutely all occupied (``filled" by the chaotic motions) and the corresponding nonlinear oscillators~(\ref{447}) must be all in dephased state, in order to permit transport to the large scales. Indeed, removing one point from a chain disconnects owing to the one-dimensionality the escape path to large wave numbers destroying the field-spreading. In the meanwhile, the Boltzmann factor in Eq.~(\ref{B+}) signifies that a space-filling distribution with $p_c = 1$ attracts the nonlinear feedback dynamics in the parameter range $s < 1$; in particular, the threshold condition $p_c = 1$ is naturally satisfied in the thermodynamic limit $\Delta n \rightarrow \infty$. 

One sees that the dephased oscillators must form kind of ``stripes" in the Hilbert space to allow transport of the wave function to large distances. Next, because the percolation regime with the integer $p_c =1$ is simultaneously critical and space-filling, the ensuing transport is just {chaotic} along the entire borderline $p_c = 1$, despite that the communication rule is next-neighbor-like. 
The onset of chaotic transport may have a high initial energy cost, though, which is essentially the cost of the stripes. As the concentration $p$ of the dephased oscillators approaches $p_c$, the percolation correlation (i.e., the pair connectedness) length diverges as $\xi \sim |p - p_c|^{-\nu}$, with $\nu = 1$ (in one dimension) \cite{Stauffer}. Thus, the nonlinear field generates via back-reaction on wave number space a long-range ordering of the percolation type, which is self-organized. We consider this ordering as favoring the field-spreading along the lattice also involving the L\'evy flights \cite{Chapter,UFN}. 

Regarding the asymptotic transport laws for the spreading, these can readily be imported from our result in Eq.~(\ref{MS-ext}), where the exponent $s$ of the power nonlinearity is now extended to the entire $2s \geq 1$. Then, as usual, $\mu =1$ for next-neighbor random walks, and $0 < \mu <1$ for nonlocal transport regimes with flights. In the local transport case, characterized by $\mu = 1$, the growth of the second moments is limited to so-called ``double diffusion" process, a jargon term designating the ubiquitous scaling 
\begin{equation}
\langle (\Delta n) ^2 (t) \rangle \propto t^{1/2}, \ \ \ t\rightarrow+\infty.
\label{DD} 
\end{equation}       
More so, adopting $\mu\rightarrow 0$, one finds that an ever achievable in terms of the moments' growth field-spreading accounting for L\'evy flights corresponds with the diffusive scaling $\langle (\Delta n) ^2 (t) \rangle \propto t$, characterized by $\alpha = 1$. We hasten to note that this ``diffusion" of the wave function is absolutely anomalous in that it arises from a competition between nonlocality of the L\'evy motion and the topological constraints contained in $s=1/2$. Indeed the inclusion of L\'evy flights introduces some nonlocality into the transport, but it does not really generate a superdiffusive scaling because of the range-dependence of the driving noise term. Hence the transport is {\it simultaneously} nonlocal and (sub)diffusive. These ``strange" transport regimes, falling off the standard picture of nonlocal behavior \cite{Klafter,Chechkin}, have been already considered in Refs. \cite{UFN,PhysScr} for separatrix dynamics in wave-like plasma turbulence.

\section{Summary and Conclusions} We consider the problem of dynamical localization-delocalization of waves in a class of nonlinear Schr\"odinger models with random potential on a lattice and arbitrary power nonlinearity. It is shown that the quadratic nonlinearity, characterized by $s=1$, plays a dynamically very distinguished role in that it is the only type of power nonlinearity to generate an abrupt localization-delocalization transition with unlimited spreading already at the delocalization border. 
We describe this localization-delocalization transition as a percolation transition on the infinite Cayley tree (Bethe lattice). The main idea here is that delocalization occurs through infinite clusters of chaotic states on a Bethe lattice, with occupancy probabilities decided by the strength of nonlinear interaction. Then the percolation transition threshold can be translated into a critical value of the nonlinearity parameter, such that above a certain critical strength of nonlinearity the field spreads to infinity, and is dynamically localized in spite of these nonlinearities otherwise. We find this critical value to be $\beta_c = 1/\ln 2 \approx 1.4427$, a fancy number representing the topology of nonlinear interaction due to the quadratic power term. 

It was argued that in vicinity of the criticality the spreading of the wave field is subdiffusive in the limit $t\rightarrow+\infty$, and that the second moments grow with time as a powerlaw $\propto t^\alpha$, with $\alpha = 1/3$ exactly. This critical regime is very special in that it stems from the direct proportionality between the nonlinear frequency shift and the distance between the excited modes in wave number space. Topologically, it corresponds with a next-neighbor random walk at the onset of percolation on a Cayley tree. The phenomena of critical spreading find their significance in some connection with the general problem of transport along separatrices of dynamical systems with many degrees of freedom \cite{ChV,EPL} and are mathematically related with a description in terms of Hamiltonian pseudochaos and time-fractional diffusion equations. 

Above the delocalization point, we find, with the criticality effects stepping aside, that the transport of the wave function turns into a chaotic domain. Yet, it is very slow (subdiffusive) involving inhomogeneity of the nonlinear interaction. The chaotic character of dynamics stipulates a Markovian diffusion equation with a range-dependent diffusion coefficient absorbing the nonhomogeneity features. The transport exponent is found to be $\alpha = 2/5$ consistently with the results of numerical simulation in Refs. \cite{Sh93,PS}. By contrast, the onset spreading ($\alpha = 1/3$) is characterized by the presence of time correlations on many scales, with algebraic auto-correlation, and is non-Markovian. We should stress that there exists a parameter range, which we identify as $\beta < \beta_c$, where the Anderson localization survives the nonlinearities. Support for this type of behavior can be found in the theoretical analyses of Refs. \cite{Fishman,Wang}. 

In case of arbitrary power nonlinearity, the patterning is to a some extent similar, but with a few important new features taking place. The general reason for the observed differences is that for $s\ne 1$ the nonlinear frequency shift is {\it not} directly proportional with the distance between the resonances in wave number space. The main points of attention consist in the following.

For a subquadratic nonlinearity, with $0 < s < 1$, the behavior is sensitive to details of definition of the nonlinear term. Employing for nonlinear interactions the power $s$ of the probability density, and trusting in Eq.~(\ref{PNL}), one generates in the Hilbert mapping space an everywhere disconnected structure, which does not permit transport in either dynamical regime (chaotic or pseudochaotic). The implication is that subquadratic nonlinearity is too weak to make it with randomness, so that the phenomena of Anderson localization occur in the nonlinear model in much the same way as in linear models \cite{And}. 

Even so, the non-analiticity of the modulus function raises concerns regarding {\it smoothness} of the mapping procedure, when the nonlinearity $|\psi_n|^{2s} \equiv \left[\psi_n\psi_n^*\right]^s$ is involved with $0 < s < 1$; hence the above conclusion that the transport does not occur for subquadratic nonlinearity turns out to be {\it not} robust in the end. Indeed defining the nonlinear term as $(\psi_n + \psi_n^*)^{2s}$  improves via the phase dependence the smoothness properties of the mapping. Then unlimited transport is, in fact, confirmed in the parameter range $1/2 \leq s<1$, and is shown to be chaotic. This behavior is mediated by one-dimensional ``stripes" of dephased oscillators in the Hilbert mapping space and is robust in the thermodynamic limit. It was argued based on the criticality of the stripy ordering that the phenomena of field-spreading were limited to ``double diffusion" (i.e., $\alpha_{\max} = 1/2$), provided just that the dynamics are local, that is, with next-neighbor jumps only; and to a diffusive scaling, with $\alpha_{\max} = 1$, for nonlocal regimes with flights. Thus, the inclusion of L\'evy flights does not really introduce a superdiffusive scaling into the modified Anderson system, so that the transport is {\it simultaneously} nonlocal and (sub)diffusive.  

It is worth mentioning here that the idea of ``stripes" has been also discussed \cite{PRB02,PLA} in connection with the phenomena of superconductivity in self-assembling complex materials \cite{Cho,Emery}, such as for instance self-assembling organic polymers and copper-oxide compounds; where it has been entitled to explain the flow of coupled Cooper pairs without resistance. Even so, the existence of superconductivity in the presence of strong underlying disorder has not been completely understood. Here, we propose based on the modified DANSE model in Eq.~(\ref{1s_Mod}) that superconductivity can survive the disorder via the nonlinear term $(\psi_n + \psi_n^*)\psi_n$ generating stripes. Then the Anderson localization of the Cooper pairs does {\it not} occur, since the nonlinear field builds by itself a channel along which it can propagate to large distances. In this regard, tiny oscillations of the stripes might also offer a natural pairing mechanism mediating the phase transition into superconducting state \cite{PRB02}. One thus predicts that superconductivity of complex materials is a synergetic wave phenomenon; where the stripy ordering, which is self-organized, does a two-fold job: (i) it generates a channel for unlimited transport of the wave function on the one hand; and (ii) provides a pairing mechanism via tiny vibrations of the chaos stripes on the other hand. 

For $s>1$, the reservations regarding smoothness and non-analyticity of the modulus function can be relaxed, implying that the nonlinear Anderson model using $|\psi_n|^{2s}$ is robust and well defined. Then likewise to the above case of subquadratic nonlinearity an unlimited transport of the wave field is only found in the chaotic regime. The behavior of the second moments is powerlaw-like involving some $s$-dependence in the transport exponent, i.e., $\alpha_s = 2 / (2s + 3)$, but with numerically smaller values as compared to the quadratic nonlinearity case. Essentially the same (i.e., chaotic) scaling is obtained for the front of diffusion on loopless fractals; where a minimal-distance the so-called chemical metric \cite{Havlin} is defined by the overlap integral in the nonlinear Anderson model. Thus, the front diffusion problem correctly phrases the chaotic regime of field-spreading in the limit $t\rightarrow+\infty$. It does {not} rephrase the pseudochaotic scaling on finite clusters; 
nor the critical spreading for $p\rightarrow p_c$ in the special case of quadratic power nonlinearity.

It was argued that the phenomena of pseudochaotic spreading driven by a superquadratic nonlinearity, $s > 1$, contained limitations on the accessible number of states. A maximal number of accessible states is controlled by the $\beta$ value and is achieved automatically through the transport. 
An important feature arising in this behavior is {\it feedback} of the spreading process on the dynamical state of the lattice, so that the expansion of the wave field is self-controlling. 
The feedback occurs via a nonlinear twist between the probability of site occupancy and the number of states already occupied by the wave field. When considered on a Bethe lattice, the spreading process is associated with self-organization to a state of critical percolation without tuning of parameters. This behavior bears signatures, enabling to classify it in terms of self-organized criticality (SOC) \cite{Bak} dynamics in the Hilbert mapping space. 

%
%

Because of feedback, pseudochaotic spreading of the nonlinear field with $s > 1$ is blocked above a certain level by transition to regular behavior. This transition occurs exactly at the percolation point, i.e., $p=p_c$, as no further spreading is allowed by the lack of connectedness. The asymptotic state of the field is a quiescent state at the border of regularity, characterized by the presence of finite clusters of dephased oscillators, with virtually no transport in the limit $t\rightarrow+\infty$. The Boltzmann factor in Eq.~(\ref{B+}) shows that a ``universal" critical strength of nonlinearity, separating the chaotic and the regular transport regimes, only exists for quadratic nonlinearity, i.e., $s=1$; where it is expressible in terms of the percolation threshold according to $p_c = \exp (-1 / \beta_c)$; and is dynamically evolving through the spreading process otherwise ($s > 1$) depending on the number of already occupied states, $\Delta n = \Delta n (t)$. In this regard, the $\beta$ value determines the running concentration of dephased oscillators in the Hilbert mapping space as a function of the running $\Delta n (t)$ value; but not really the fact that the field does or does not spread to infinity in wave number space.

Other than providing a connection to self-controlling dynamics and SOC, nonlinear Schr\"odinger models with randomness offer a fertile environment for the generalized kinetic equations built on fractional derivative operators. In the above we have encountered two types of such equations. One type is associated with a fractional extension of the time derivative; this type of fractional diffusion occurs in models of pseudochaotic transport along separatrices of spatially extended systems \cite{EPL,PRE09,NJP}. The second type is associated with chaotic dynamics in the presence of a competing nonlocal ordering \cite{PRB02,PLA,UFN}; a L\'evy-fractional diffusion equation, involving fractional extension of the Laplacian, is a prominent example of this type \cite{Klafter,Chechkin,Jesper}. It is noticed that no fractional extension of the original DANSE has been assumed to obtain these fractional equations. This observation also emphasizes the different physics implications behind the fractional kinetic {\it vs.} dynamical equations \cite{Chapter,IominFT,Comb}. 

In the discussion above it was argued that the interaction between the components of the wave field leads to a range-dependent diffusion model in the parameter range of chaotic transport, implying that the diffusion coefficient scales with the number of states. This dependence when cast into the general framework of the L\'evy statistics has an important impact onto the nonlocal behavior. It leads to a competition between the nonlocality of the L\'evy motion and the range-dependence of the diffusion coefficient of the waves. As a result, the behavior is subdiffusive in spite of L\'evy flights. 

More so, there is an upper bound on the rate of nonlocal spreading, which is found in the standard model~(\ref{1s}) at the margins of stability of the motion of the L\'evy type in the limit $s\rightarrow 1$ from above. It corresponds to a subdiffusive process with the exponent $\alpha_{\max} = 2/3$. We consider this transport regime as a theoretical prediction of our approach. A generalization of this behavior to a DANSE model with modified nonlinearity, Eq.~(\ref{1s_Mod}), have been also addressed, leading to the upper bound $\alpha_{\max} = 1$ instead. 
A summary of transport exponents for the various asymptotic regimes of field-spreading in the standard DANSE~(\ref{1s}) is collected in Table~1. 

\begin{table}[t]
\caption{Transport exponents for the various regimes of field-spreading in the standard DANSE model: Eq.~(\ref{1s}).} \label{tab1}
\begin{center}
\begin{tabular}{p{3.9cm}p{2.0cm}p{2.3cm}} \hline \hline
Regime & $s=1 $ & $s > 1 $  \\ \hline 
Chaotic (Gaussian) & $2/5$ & $2/(2s+3)$ \\ 
Chaotic (L\'evy) & $2/(2\mu+3)$ & $2 / (2\mu+2s+1)$\footnote{$\max (\alpha)$ = $2/3$ for the standard DANSE model: Eq.~(\ref{1s})}\footnote{$\max (\alpha)$ = $1$ for the modified DANSE model: Eq.~(\ref{1s_Mod})} \\ 
\hline
Pseudochaotic ($p\rightarrow p_c$) & $1/3$ & $1/(2s + 1)$\footnote{For $1\ll t\ll (\Delta n_{\max})^{2(2s+1)}$} \\ 
\hline \hline
\end{tabular}
\end{center}
\label{default}
\end{table}

The main emphasis in the present work has been laid on power nonlinearity, thought as a suitable toy model in describing the properties of interaction between the components of the wave field. Generalizations of this correspond to analytical nonlinearity, given by a series expansion of the probability density. One prospective model here with a nontrivial phase transition-like behavior is defined by the equation 
\begin{equation}
i\hbar\frac{\partial\psi_n}{\partial t} = \hat{H}_L\psi_n + \left[e^{\beta |\psi_n|^{2s}} - 1\right] \psi_n,
\label{Exp} 
\end{equation}
where $s$ ($s > 0$) is the parameter of the interaction. Expanding the exponential function in powers of $|\psi_n|^{2s}$, one can immediately become convinced that the models with $s = 1/n$, where $n = 1, 2,\dots$ is a natural number, will include a quadratic nonlinearity; which, therefore, will be responsible for a localization-delocalization transition, and for unlimited spreading of the wave field both at and above the delocalization border. The threshold of delocalization is found to be 
\begin{equation}
\beta_c = \left[\Gamma (1 + 1/s) / \ln 2\right]^s, 
\label{Thr} 
\end{equation}
where Eq.~(\ref{Betac}) has been considered. So, as the strength of nonlinearity $\beta$ approaches the critical strength $\beta_c$, the field abruptly turns into a delocalized state, giving rise to an unlimited spreading of the wave function along the lattice. We might predict that in vicinity of the criticality the spreading is subdiffusive for $t\rightarrow+\infty$, i.e., $\langle (\Delta n) ^2 (t) \rangle \propto t^{\alpha}$, and that the exponent $\alpha = 1/3$ (same as in models with pure quadratic nonlinearity). 

Although obvious, it should be emphasized that the unlimited spreading, along with a phase transition-like behavior, only exists for the rational values $s = 1,\, ^1/_2,\, ^1/_3, \dots$, which guarantee the presence of the quadratic term in the expansion. One sees that the quadratic nonlinearity is instrumental in describing the critical regimes of wave-spreading in seemingly very different theoretical models. 

It is worth noting that delocalization by nonlinear interaction is a mechanism of sufficiently general nature. As such, it may be extended so that it includes phenomena beyond the strict context of the Anderson problem as for instance phenomena of self-delocalization of fractons \cite{Chapter,UFN}, beam-plasma systems in a toroidal geometry \cite{NF2005,PPCF06,Heid}, and the Fock-space delocalization problem \cite{Alt}. In this respect, we should stress that theoretical investigations, presented here, are the basis for consistency analysis of the different localization-delocalization patterns in systems with many interacting degrees of freedom in association with the asymptotic properties of the transport.

\acknowledgments
A.V.M. and A.I. thank the Max-Planck-Institute for the Physics of Complex Systems (Dresden, Germany) for hospitality and financial support. This work was supported in part 
by the Israel Science Foundation (ISF) and by the ISSI project ``Self-Organized Criticality and Turbulence" (Bern, Switzerland).


\begin{thebibliography}{}

\bibitem{Christ}
P. L. Christiansen, Yu. B. Gaididei, K. \O. Rasmussen, V. K. Mezentzev, and J. Juul Rasmussen, Phys. Rev. B.  {\bf 54}, 900 (1996).

\bibitem{Gaid}
P. L. Christiansen, Yu. B. Gaididei, M. Johansson, K. \O. Rasmussen, V. K. Mezentzev, and J. Juul Rasmussen, Phys. Rev. B.  {\bf 57}, 11 303 (1998).

\bibitem{Clement}
D. Cl\'ement, A. F. Var\'on, M. Hugbart, J. A. Retter, P. Bouyer, L. Sanchez-Palencia, D. M. Gangardt, G. V. Shlyapnikov, and A. Aspect, Phys. Rev. Lett. {\bf 95}, 170409 (2005).

\bibitem{Shapiro}
B. Shapiro, Phys. Rev. Lett. {\bf 99}, 060602 (2007).

\bibitem{Blumen}
H. Buttner and A. Blumen, Nature (London) {\bf 329}, 700 (1987).

\bibitem{PRB02}
A. V. Milovanov and J. Juul Rasmussen, Phys. Rev. B {\bf 66}, 134505 (2002).

\bibitem{Alt}
B. L. Altshuler, Y. Gefen, A. Kamenev, and L. S. Levitov, Phys. Rev. Lett. {\bf 78}, 2803 (1997).

\bibitem{Chapter}
A. V. Milovanov, in {\it Self-Organized Criticality Systems} (Ed. M. J. Aschwanden, Open Academic Press, Berlin, Warsaw, 2013) ({\it Percolation Models of Self-Organized Critical Phenomena}, Chapter 4, pp. 103-182). 

\bibitem{Bak}
P. Bak, C. Tang, and K. Wiesenfeld, Phys. Rev. Lett. {\bf 59}, 381 (1987); Phys. Rev. A {\bf 38}, 364 (1988).

\bibitem{And}
P. W. Anderson, Phys. Rev. {\bf 109}, 1492 (1958).

\bibitem{Electron}
E. Akkermans and G. Montambaux, {\it Mesoscopic Physics of Electrons and Photons} (Cambridge Univ. Press, Cambridge, 2006).

\bibitem{Sound}
R. L. Weaver, Wave Motion {\bf 12}, 129 (1990).

\bibitem{Maret}
M. St\~orzer, P. Gross, C. M. Aegerter, and G. Maret, Phys. Rev. Lett. {\bf 96}, 063904 (2006).

\bibitem{Light}
T. Schwartz, G. Bartal, S. Fishman, and M. Segev, Nature (London) {\bf 446}, 52 (2007). 

\bibitem{BE}
J. Billy, V. Josse, Z. Zuo, A. Bernard, B. Hambrecht, P. Lugan, D. Cl\'ement, L. Sanchez-Palencia, P. Bouyer, and A. Aspect, Nature (London) {\bf 453}, 891 (2008).

\bibitem{Sh93}
D. L. Shepelyansky, Phys. Rev. Lett. {\bf 70}, 1787 (1993).

\bibitem{PS}
A. S. Pikovsky and D. L. Shepelyansky, Phys. Rev. Lett. {\bf 100}, 094101 (2008). 

\bibitem{Flach}
S. Flach, D. O. Krimer, and Ch. Skokos, Phys. Rev. Lett. {\bf 102}, 024101 (2009).

\bibitem{Skokos}
Ch. Skokos, D. O. Krimer, S. Komineas, and S. Flach, Phys. Rev. E {\bf 79}, 056211 (2009).

\bibitem{EPL}
A. V. Milovanov and A. Iomin, Europhys. Lett. {\bf 100}, 10006 (2012). 

\bibitem{Gross}
F. Dalfovo, S. Giorgini, L. P. Pitaevskii, and S. Strigani, Rev. Mod. Phys. {\bf 71}, 463 (1999).

\bibitem{Edros}
L. Erd\"os, B. Schlein, and H. T. Yau, Phys. Rev. Lett. {\bf 98}, 040404 (2007).

\bibitem{Sagdeev}
G. M. Zaslavsky and R. Z. Sagdeev, {\it Introduction to the Nonlinear Physics. From Pendulum to Turbulence and Chaos} (Nauka, Moscow, 1988). 

\bibitem{Report}
G. M. Zaslavsky, Phys. Rep. {\bf 371},  461 (2002).

\bibitem{Klafter}
R. Metzler and J. Klafter, Phys. Rep. {\bf 339}, 1 (2000). 

\bibitem{Chechkin}
R. Metzler, A. V. Chechkin, V. Yu. Gonchar, and J. Klafter, Chaos, Solitons \& Fractals {\bf 34}, 129 (2007).

\bibitem{Arnold}
V. I. Arnold, {\it Mathematical Methods of Classical Mechanics} (Springer, Berlin, 1978).

\bibitem{ChV}
B. V. Chirikov and V. V. Vecheslavov, Zh. \'Eksp. Teor. Fiz. {\bf 112}, 1132 (1997).

\bibitem{JMPB}
O. Lyubomudrov, M. Edelman, and G. M. Zaslavsky, Intl. J. Mod. Phys. B {\bf 17}, 4149 (2003).

\bibitem{PhysD}
G. M. Zaslavsky and M. A. Edelman, Physica D {\bf 193}, 128 (2004).

\bibitem{PRE97}
A. V. Milovanov, Phys. Rev. E {\bf 56}, 2437 (1997).

\bibitem{PRE09}
A. V. Milovanov, Phys. Rev. E {\bf 79}, 046403 (2009). 

\bibitem{NJP}
A. V. Milovanov, Europhys. Lett. {\bf 89}, 60004 (2010); New J. Phys. {\bf 13}, 043034 (2011). 

\bibitem{ZaslavskyUFN}
G. M. Zaslavsky and B. V. Chirikov, Phys. Usp. {\bf 14}, 549 (1972); G. M. Zaslavsky, {\it Statistical Irreversibility in Nonlinear Systems} (Nauka, Moscow, 1970).

\bibitem{Havlin}
S. Havlin and D. ben-Avraham, Adv. Phys. {\bf 51}, 187 (2002); D. ben-Avraham and S. Havlin, {\it Diffusion and Reactions in Fractals and Disordered Systems} (Cambridge University Press, Cambridge, 2000).


\bibitem{Van}
N. G. van Kampen, {\it Stochastic Processes in Physics and Chemistry} (North-Holland, Amsterdam, 1981).


\bibitem{Gnedenko}
B. V. Gnedenko and A. N. Kolmogorov, {\it Limit Distributions for Sums of Independent Random Variables} (Addison-Wesley, Reading, 1954).

\bibitem{PLA}
A. V. Milovanov and J. Juul Rasmussen, Phys. Lett. A {\bf 337}, 75 (2005).

\bibitem{Cho}
A. Cho, Phys. Rev. Focus {\bf 9} (2002).

\bibitem{UFN}
L. M. Zelenyi and A. V. Milovanov, Physics-Uspekhi {\bf 47}, 749 (2004). 

\bibitem{JJR}
A. V. Milovanov and J. Juul Rasmussen, Phys. Lett. A {\bf 378}, 1492 (2014); 
arXiv:1403.5896v1 [nlin.CD].

\bibitem{Emery}
S. A. Kivelson, E. Fradkin, and V. J. Emery, Nature (London) {\bf 393}, 550 (1998); E. W. Carlson, D. Orgad, S. A. Kivelson, and V. J. Emery, Phys. Rev. B {\bf 62}, 3422 (2000).

\bibitem{And+}
R. Abou-Chacra, P. W. Anderson, and D. J. Thouless, J. Phys. C: Solid State Phys. {\bf 6}, 1734 (1973).

\bibitem{Hirsch}
M. W. Hirsch, {\it Differential Topology} (Springer-Verlag, New York, 1976).

\bibitem{Chirikov}
B. V. Chirikov, J. Nucl. Energy Part C: Plasma Phys. {\bf 1}, 253 (1960).

\bibitem{PRE01}
A. V. Milovanov, Phys. Rev. E {\bf 63}, 047301 (2001). 

\bibitem{PRE00}
A. V. Milovanov and G. Zimbardo, Phys. Rev. E {\bf 62}, 250 (2000). 

\bibitem{Isi}
M. B. Isichenko, Rev. Mod. Phys. {\bf 64}, 961 (1992).

\bibitem{Kubo}
R. Kubo, J. Math. Phys. {\bf 4}, 174 (1963).

\bibitem{Schroeder}
M. R. Schroeder, {\it Fractals, Chaos, Power Laws: Minutes from an Infinite Paradise} (Freeman, New York, 1991).

\bibitem{Gefen}
Y. Gefen, A. Aharony, and S. Alexander, Phys. Rev. Lett. {\bf 50}, 77 (1983).

\bibitem{Stauffer}
D. Stauffer, Phys. Rep. {\bf 54}, 1 (1979); D. Stauffer and A. Aharony, {\it Introduction to Percolation Theory} (Taylor \& Francis, London, 1992).

\bibitem{AO}
S. Alexander and R. Orbach, J. Phys. Lett. (Paris) {\bf 43}, L625 (1982).

\bibitem{Rammal}
R. Rammal and G. Toulouse, J. Phys. Lett. (Paris) {\bf 44}, L13 (1983).

\bibitem{Bouchaud}
J.-P. Bouchaud and A. Georges, Phys. Rep. {\bf 195}, 127 (1990).

\bibitem{Shaugh}
B. O'Shaughnessy and I. Procaccia, Phys. Rev. Lett. {\bf 54}, 455 (1985).

\bibitem{Naka}
T. Nakayama, K. Yakubo, and R. L. Orbach, Rev. Mod. Phys. {\bf 66}, 381 (1994).

\bibitem{Coniglio}
A. Coniglio, J. Phys. A {\bf 15}, 3829 (1982).

\bibitem{Podlubny}
I. Podlubny, {\it Fractional Differential Equations} (Academic Press, San Diego, 1999). 

\bibitem{Uch}
V. V. Uchaikin, Physics-Uspekhi {\bf 46}, 821 (2003).

\bibitem{Nature}
M. F. Shlesinger, G. M. Zaslavsky and J. Klafter, {Nature} (London) {\bf 363}, 31 (1993); I. M. Sokolov, J. Klafter, and A. Blumen, Phys. Today {\bf 55}, 48 (2002). 

\bibitem{Rest}
R. Metzler and J. Klafter, J. Phys. A: Math. Gen. {\bf 37}, R161 (2004).

\bibitem{Iomin}
A. Iomin, Phys. Rev. E {\bf 81}, 017601 (2010).

\bibitem{Ctrw1}
R. Metzler, J. Klafter, and I. M. Sokolov, Phys. Rev. E {\bf 58}, 1621 (1998).

\bibitem{Cox}
D. Cox, J. Little, and D. O'Shea, {\it Using Algebraic Geometry} (Springer-Verlag, New York, 1998).


\bibitem{Levy}
P. L\'evy, {\it Th\'eorie de l'Addition des Variables Aleatoires} (Gauthiers-Villars, Paris, 1937).

\bibitem{Barkai}
R. Metzler, E. Barkai, and J. Klafter, Europhys. Lett. {\bf 46}, 431 (1999).

\bibitem{Oldham}
K. B. Oldham and J. Spanier, {\it The Fractional Calculus} (Academic Press, New York, 1974).

\bibitem{Samko}
S. G. Samko, A. A. Kilbas, and O. I. Marichev, {\it Fractional Integrals and Derivatives. Theory and Applications} (Gordon and Breach, Amsterdam, 1993). 

\bibitem{Fog}
H. C. Fogedby, Phys. Rev. E {\bf 50}, 1657 (1994).

\bibitem{Jesper}
S. Jespersen, R. Metzler, and H. C. Fogedby, Phys. Rev. E {\bf 59}, 2736 (1999).

\bibitem{Gonchar}
A. V. Chechkin and V. Yu. Gonchar, J. Exp. Theor. Phys. {\bf 91}, 635 (2000).

\bibitem{Srok}
T. Srokowski, Phys. Rev. E {\bf 79}, 040104 (2009).

\bibitem{Levich}
V. G. Levich, {\it Course of Theoretical Physics. Vol. 1} (Fizmatgiz, Moscow, 1962).

\bibitem{JSTAT}
A. V. Milovanov and A. Iomin, J. Stat. Mech. (submitted, 2014); arXiv: 1405.7510v1 [cond-mat.stat-mech].

\bibitem{Skokos_PRE}
Ch. Skokos and S. Flach, Phys. Rev. E {\bf 82}, 016208 (2010). 

\bibitem{Ash2013}
M. J. Aschwanden, Ed. {\it Self-Organized Criticality Systems} (Open Academic Press GmbH \& Co., 2013).

\bibitem{SSRv}
M. J. Aschwanden, N. Crosby, M. Dimitropoulou, M. K. Georgoulis, S. Hergarten, H.J. Jensen, J. McAteer, A. V. Milovanov, S. Mineshige, L. Morales, N. Nishizuka, G. Pruessner, R. Sanchez, S. Sharma, A. Strugarek, and V. Uritsky, Space Sci. Rev. (accepted, 2014); arXiv:1403.6528v1 [astro-ph.IM].

\bibitem{PhysScr}
F. Chiaravalloti, A. V. Milovanov, and G. Zimbardo, Phys. Scr. {\bf T122}, 79 (2006). 

\bibitem{Fishman}
Y. Krivolapov, S. Fishman, and A. Soffer, New J. Phys. {\bf 12}, 063035 (2010).

\bibitem{Wang}
W.-M. Wang and Z. Zhang, J. Stat. Phys. {\bf 134}, 953 (2009).



\bibitem{IominFT}
A. Iomin, Phys. Rev. E {\bf 80}, 022103 (2009).

\bibitem{Comb}
A. Iomin, Chaos, Solitons $\&$ Fractals {\bf 44}, 348 (2011).

\bibitem{NF2005}
L. Chen and F. Zonca, Nucl. Fusion {\bf 45}, 477 (2005); 
F. Zonca, S. Briguglio, L. Chen, G. Fogaccia, and G. Vlad, Nucl. Fusion {\bf 45}, 477 (2005).

\bibitem{PPCF06}
F. Zonca, S. Briguglio, L. Chen, G. Fogaccia, T. S. Hahm, A. V. Milovanov, and G. Vlad, Plasma Phys. Control. Fusion {\bf 48}, B15 (2006).

\bibitem{Heid}
W. W. Heidbrink, Phys. Plasmas {\bf 15}, 055501 (2008).


\end{thebibliography}

\end{document}